# Analysis of self-equilibrated networks through cellular modeling

O. ALOUI[a], D. ORDEN[b], N. BEL HADJ ALI[c,d], L. RHODE-BARBARIGOS[a]*

[a] Department of Civil, Architectural and Environmental Engineering, University of Miami, Coral Gables, FL 33146, USA

[b] Departamento de Física y Matemáticas, Universidad de Alcalá. Ctra. Madrid-Barcelona, Km. 33,600, 28805, Alcalá de Henares, Spain

[c] Ecole Nationale d'Ingénieurs de Gabès, University of Gabès, Rue Omar Ibn-Elkhattab 6029, Gabès, Tunisia

[d] LASMAP, Ecole Polytechnique de Tunisie, University of Carthage, B.P. 743, La Marsa 2078, Tunisia

**Abstract**

Network equilibrium models represent a versatile tool for the analysis of interconnected objects and their relationships. They have been widely employed in both science and engineering to study the behavior of complex systems under various conditions, including external perturbations and damage. In this paper, network equilibrium models are revisited through graph-theory laws and attributes with special focus on systems that can sustain equilibrium in the absence of external perturbations (self-equilibrium). A new approach for the analysis of self-equilibrated networks is proposed; they are modeled as a collection of cells, predefined elementary network units that have been mathematically shown to compose any self-equilibrated network. Consequently, the equilibrium state of complex self-equilibrated systems can be obtained through the study of individual cell equilibria and their interactions. A series of examples that highlight the flexibility of network equilibrium models are included in the paper. The examples attest how the proposed approach, which combines topological as well as geometrical considerations, can be used to decipher the state of complex systems.

## 1. Introduction

Since their inception, networks have been serving as a powerful tool to model a wide range of engineering problems. In 1736, Leonhard Euler used graph representations to prove that the problem of the seven bridges of Königsberg has no solution [1]. This laid the ground for the emergence of network theory and predefined the concept of topology. However, network applications did not remain confined to the study of topological properties of systems but quickly evolved to incorporate a comprehensive description of their equilibrium states. The integration of equilibrium in networks led to the discovery of network equilibrium models. This turned out to be very influential in the study of electrical networks especially with the advances generated by the work of Kirchhoff on the "node and mesh" rules for electrical circuits [2-3]. Network equilibrium models have also been used in the analysis of mechanical and structural systems. In 1864, Maxwell formulated the counting conditions to determine the rigidity of bar-jointed frameworks based on the topology of the underlying network [4]. Maxwell's work was further refined by Calladine, Pellegrino, Roth, Whitely and Connelly [5-9], where the network structure of the framework described the equilibrium between conjugate variables: forces and displacements. These conjugate quantities are attributed to the network nodes and edges that have to satisfy nodal equilibrium and geometric compatibility. Moreover, the analogy between the analysis of electrical networks and mechanical systems was also recognized as nodal equilibrium and geometric compatibility relations in structural frameworks are reflected in Kirchhoff current and voltage laws. Hähnle and Firestone provided a complete set of analogies between electrical and mechanical systems where forces are treated as currents and displacements are considered as voltages, allowing researchers to explain electrical phenomena by referring to mechanical systems and vice-versa [10,11]. Network equilibrium models have also been adopted in the analysis of electrical and power networks [12-14], telecommunication networks [15-17], transportation networks [18-20], as well as supply chain networks [21-23].

The wide range of physical and engineering systems that are depicted through network models underlines the value of using network representation to model such systems, as the variables involved in the related problems can be attributed to the network components (nodes and edges) and the relations between these variables can be described. It is thus widely recognized that a significant portion of physical, engineering and mathematical problems lie within the scope of network theory [24-28]. However, there is often dim interest in studying the abstract topological and algebraic properties of network equilibrium models when the focus is on specific context. Nevertheless, understanding the abstract properties of network equilibrium models is critical for their application, as it provides a platform for studying interdependent models as well as a common language for interdisciplinary collaborations. Analogies can thus be drawn making possible to adopt solutions already developed in other fields. One can find some interest on the abstract properties of network problems in the work of Roth, who applied algebraic topology concepts to study the existence of a solution to the network equilibrium problem [29-31]. Branin built upon Roth's work to study the topological structure of the network or the linear graph and the associated algebraic structure, setting up ground rules for network analogies and discussing the interpretations of these rules in electrical, mechanical, and structural systems [32,33]. More recently, Reinschke provided a comprehensive description of the network equilibrium models, starting from an abstract model for the variable attributes of the network components and the network elements relations (NER) between them [34].

This paper extends Reinschke's work on network equilibrium models by focusing on a special class of network models, referred to as the self-equilibrated network models, with a novel approach for the analysis of their topological and algebraic properties based on elementary units called cells. Self-equilibrated network models present a redundancy in their elements that can be explored in science and engineering applications that require a certain degree of damage tolerance. The paper is organized as follows: Section 2 includes a review of network equilibrium models through the description of their network laws and attributes (for more details, see appendix A), as well as of their properties and equilibrium state. Self-equilibrated networks are described in Section 3 through the definition of their constitutive cells and their interactions, as well as their impact in the network attributes. In Section 4, the equilibrium in examples of self-equilibrated network models is studied considering also the effects of external perturbation and damage (element loss). Section 5 concludes the paper with a discussion for the use of the proposed model.

## 2. Network equilibrium models

Let $G(V, E)$ be a graph that describes the set of nodes $V$ and the set of edges $E$ of a network and $n_v$ be the number of vertices and $n_e$ the number of edges in the graph. The graph is equipped with a set of node and edge attributes that satisfy the equilibrium conditions referred to, in graph theory, as circuit laws and cut-set laws. In this paper, node attributes are referred to as potential attributes; they are denoted by a $n_v \times 1$ vector $p$ and they satisfy circuit laws. Note that each component $p_i$ represents the value of the potential at node $v_i$ which in turn is a $d \times 1$ vector where $d$ represents the dimensionality of the problem. Edge attributes are referred to as flow attributes; they are denoted by a $n_e \times 1$ vector $f$ and they satisfy cut-set laws. The properties of edge attributes and node attributes along with the description of cut-set laws and circuit laws are discussed in detail in Appendix A.

### 2.1. Network equilibrium

In this section, network equilibrium is described in terms of the topology of the graph and its different attributes. A network equilibrium model is thus given by the flow and potential attributes, and the interrelations between them that can be expressed by [34]:

$$r(f, p) = 0 \tag{1}$$

where $f$ is the $n_e \times 1$ vector (indexed by the edges of $G$) of the values that the flow takes on each edge and $p$ represents the $n_v \times 1$ vector (indexed by the nodes of $G$) of the values that the potential takes on the nodes. Equation 1 can thus be used to model voltage-current relations in electrical circuits or constitutive relations, such as force-displacement relations, in solid mechanics. In constitutive relations, flow and potential are related through the impedance of the electrical circuit branch (inductance, capacitance, resistance, etc.) or the stiffness of the structural member. Similar concepts can be found in transportation networks. Figure 1 represents a network equilibrium model with edge and end-nodes equipped with flow and potential attributes.

In Figure 1, $u$ and $v$ are the end-nodes of the edge $(u, v)$. Nodes $u$ and $v$ are attributed an intrinsic potential represented by the value the potential function $p$ takes on $u$ and $v$ and an independent potential $p^e$. This incurs the potential difference variable $\delta p((u,v))$ on the edge $(u,v)$ and $\delta p^e$. Edge $(u, v)$ is also attributed the intrinsic flow represented by the value of the flow function $f$ on $(u, v)$ and the independent flow $f^e$. Independent flow variables can be interpreted as independent current source in electrical circuits or perturbations to the element stresses in structural members such as thermal expansion. In transportation networks, $f^e$ can be used to model perturbation to the flow of goods due to external agents such as a change in edge capacity. The independent potential variable accounts for independent voltage sources in electrical circuits, external loads or displacements applied to the nodes of a structure or changes in the stock due to creation of new quantities of goods and/or additional traveling agents in transportation systems. $r(f, p)$ represents the flow-potential relations.

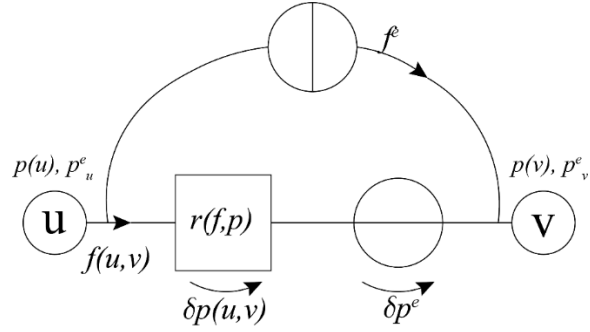

Figure 1: Edge and end-node attribute representation in a network equilibrium model. $p(u)$: potential values in nodes $u, v$. $\delta p(u,v)$: potential difference values between nodes $u$ and $v$. $p_u^e$: independent potential variable in node $u$. $\delta p^e$: independent potential difference value between nodes $u$ and $v$. $f(u,v)$: flow values on edge $(u,v)$. $f^e$: independent flow value on edge $(u,v)$

Since the flow space $\mathcal{F}$ and the potential difference space $\delta\mathcal{P}$ are orthogonal complements of each other, and $\mathcal{F}$ and $\delta\mathcal{P}$ are associated to the cycle space $\mathcal{C}$ and the cut-set space (bond space) $\mathcal{B}$, for the determination of whether an edge attribute $f$ is a flow it suffices to verify its orthogonality with a basis of the cut-set space $\mathcal{B}$. Let $B$ be the matrix formed by vectors $(b_1, b_2, \ldots, b_r)^T$, then the cut law can be expressed in matrix form as:

$$B(f + f^e) = 0 \tag{2}$$

Analogously the circuit law can be expressed as:

$$C(\delta p + \delta p^e) = 0 \tag{3}$$

where $C$ is the collection of the cycle space bases. The network equilibrium model in a one-dimensional space is thus described by:

$$\begin{cases} B(f + f^e) = 0 & \text{cutset laws} \\ C(\delta p + \delta p^e) = 0 & \text{circuit laws} \\ r(f + f^e, p + p^e) = 0 & \text{flow and potential relations} \end{cases} \tag{4}$$

2.2. Network equilibrium models in higher dimensions

Networks that are embedded in higher dimensions can be used to describe a wide range of physical and engineering systems as well as model one-dimensional systems with interdependencies. Therefore, this section discusses the dimensionality of the network equilibrium models. The dimensionality of the flow variables can be described as interconnected networks of the same topology and same embedding (potentials) with each network being endowed with a one-dimensional component of flow and flow-potential relations that describe the inter-dependence of flow components. Conversely, the dimensionality of the potential variables introduces the concept of direction (described by vectors) to the network models. In this paper, directions refer to the generalization of the concept of orientation in higher dimensions. For instance, in DC electrical circuits (which correspond to one-dimensional networks), flow orientation is set by convention from higher voltage to lower voltage (negative potential difference). In structural systems, direction is set by a position vector given by a unit vector obtained by dividing the "potential difference" by the length of the edge. Consequently, direction is an inherent part of the description of the network equilibrium. Consider an elementary cut $\Delta_u = [\{u\}, V/\{u\}]$ in the network $G(V, E)$ equipped with a potential $p$ and a flow $f$. The cut law is given by:

$$\sum_{\substack{v \in V \\ (u,v) \in E}} f((u,v)) = 0 \tag{5}$$

Note that in Equation 5, the direction is already incorporated in the flow $f(u, v)$ described by $f(u, v) = \|f(u, v)\| . \vec{e}_{(u,v)}$ as $\|f(u, v)\|$ represents the magnitude of the flow and $\vec{e}_{(u,v)}$ corresponds to a unit vector that depicts the direction of the edge $(u, v)$. Now, let $d(u, v)$ be a distance function between nodes $u$ and $v$. In one dimension, the distance is defined as $d(u, v) = |p(u) - p(v)| = \|\delta p((u, v))\|$. In higher dimensions, the distance $d$ is recognized as the Euclidean distance (L2 norm) where $d(u, v) = \|\delta p((u, v))\|^2$. In two dimensions, $d(u, v) = \sqrt{\delta p(u, v)_x^2 + \delta p(u, v)_y^2}$. The flow density $\omega^f(u, v)$ is defined as the quantity $\frac{\|f(u,v)\|}{\|\delta p(u,v)\|}$. By introducing the flow density and the distance function, the cut laws can be expressed as:

$$\sum_{\substack{v \in V \\ (u,v) \in E}} f((u,v)) = 0$$
$$\Leftrightarrow \sum_{\substack{v \in V \\ (u,v) \in E}} \|f((u,v))\| . \vec{e}_{(u,v)} = 0$$
$$\Leftrightarrow \sum_{\substack{v \in V \\ (u,v) \in E}} \|f((u,v))\| . \frac{\delta p((u,v))}{\|\delta p((u,v))\|} = 0 \tag{6}$$
$$\Leftrightarrow \sum_{\substack{v \in V \\ (u,v) \in E}} \omega^f((u,v)) . \delta p((u,v)) = 0$$
$$\Leftrightarrow \sum_{\substack{v \in V \\ (u,v) \in E}} \omega^f((u,v)) . (p(u) - p(v)) = 0$$

Using the flow density allows to redefine the equilibrium problem in terms of a scalar quantity reducing the number of variables. A multidimensional network problem can thus be simplified as interrelated networks of the same topology that share the same flow density. When all elementary cut-sets are considered, the cut laws can be described in matrix form as:

$$A^{\delta p} \cdot (\omega^f + \omega^{fe}) = 0$$

$$\Leftrightarrow \begin{pmatrix} B \; diag(\delta p^{x_1}) \\ B \; diag(\delta p^{x_2}) \\ \vdots \\ B \; diag(\delta p^{x_d}) \end{pmatrix} (\omega^f + \omega^{fe}) = 0 \quad (7)$$

$$\Leftrightarrow (\mathbb{I}_d \otimes B) \cdot \begin{pmatrix} diag(\delta p^{x_1}) \\ diag(\delta p^{x_2}) \\ \vdots \\ diag(\delta p^{x_d}) \end{pmatrix} (\omega^f + \omega^{fe}) = 0$$

$$\Leftrightarrow (\mathbb{I}_d \otimes B) \cdot diag(\delta p) \cdot (J_{d,1} \otimes \mathbb{I}_m)(\omega^f + \omega^{fe}) = 0$$

where $J_{d,1}$ is an all-ones $d \times 1$ vector, and $\delta p$ is the $dn_e \times 1$ vector of potential differences, ordered such that all the components of the same dimension are grouped together. These components are denoted $\delta p^{x_1}$, $\delta p^{x_2}$, ..., $\delta p^{x_d}$. Note that $\delta p$ can be expressed as $\delta p = (\mathbb{I}_d \otimes B^T) \cdot p$, where $p$ is the vector of potential values where the components of each dimensions $X_1, X_2, \ldots, X_d$ are grouped together. $\mathbb{I}_d$ is the $d \times d$ identity matrix. $B$ is the $n_v \times n_e$ matrix that groups all the elementary cut-sets. $\omega^f$ and $\omega^{fe}$ are $n_e \times 1$ vectors of the internal flow densities and independent flow densities respectively. $diag(u)$ is the function that takes a $1 \times n$ vector $u$ and returns a $n \times n$ diagonal matrix $U$. The matrix $A^{\delta p}$ is known in the analysis of pin-jointed frameworks as the equilibrium matrix, where a more comprehensible form can be expressed as:

$$A^{\delta p} = \begin{pmatrix} B \; diag(B^T X_1) \\ B \; diag(B^T X_2) \\ \vdots \\ B \; diag(B^T X_d) \end{pmatrix} \quad (8)$$

with $(X_1, X_2, \ldots, X_d)$ being the components of each dimension of $p$. Note that this form contains redundant rows. The redundancy created by the topology of the graph can be omitted by using $B_r$ instead of $B$, where the rows represent the basis of the cut-set space. However, the dimensionality of the problem might also create redundant rows.

Circuit laws can also be described based on the potential density $\omega^{\delta p}(u,v) = \frac{\|\delta p(u,v)\|}{\|f(u,v)\|}$, with the circuit law for a given cycle $C$ becoming:

$$\sum_{(u,v) \in C} \delta p\big((u,v)\big) = 0$$

$$\Leftrightarrow \sum_{(u,v) \in C} \|\delta p\big((u,v)\big)\| \cdot \vec{e}_{(u,v)} = 0$$

$$\Leftrightarrow \sum_{(u,v) \in C} \|\delta p\big((u,v)\big)\| \cdot \frac{f\big((u,v)\big)}{\|f\big((u,v)\big)\|} = 0 \quad (9)$$

$$\Leftrightarrow \sum_{(u,v) \in C} \omega^{\delta p}\big((u,v)\big) \cdot f\big((u,v)\big) = 0$$

Considering all independent cycles of the graph, the circuit law can be expressed in matrix form as:

$$\begin{aligned} & A^f \cdot \left( \omega^{\delta p} + \omega^{\delta p^e} \right) = 0 \\ \Leftrightarrow & \begin{pmatrix} C\, diag(f^{x1}) \\ C\, diag(f^{x2}) \\ \vdots \\ C\, diag(f^{xd}) \end{pmatrix} \left( \omega^{\delta p} + \omega^{\delta p^e} \right) = 0 \\ \Leftrightarrow & (\mathbb{I}_d \otimes C) \cdot \begin{pmatrix} diag(f^{x1}) \\ diag(f^{x2}) \\ \vdots \\ diag(f^{xd}) \end{pmatrix} \left( \omega^{\delta p} + \omega^{\delta p^e} \right) = 0 \\ \Leftrightarrow & (\mathbb{I}_d \otimes C) \cdot diag(f) \cdot (J_{d,1} \otimes \mathbb{I}_m)\left( \omega^{\delta p} + \omega^{\delta p^e} \right) = 0 \end{aligned} \qquad (10)$$

where $f$ is the $dn_e \times 1$ vector of flow values, ordered so that all components of the same dimension are grouped together. $A^{\delta p}$ and $A^f$ define the static equilibrium of the model. In other words, if the system is in equilibrium, Equations 7 and 10 have to be satisfied. This allows finding the equilibrium flows in a network when the potential is known, and vice versa. When a perturbation occurs to the flows or the potentials, the new equilibrium is governed by the network element relation $r(f + f^e, p + p^e) = 0$ which can be expressed using the flow and potential densities as $r(\omega^f + \omega^{f^e}, \omega^{\delta p} + \omega^{\delta p^e}) = 0$ or $r(\omega^f + \omega^{f^e}, p + p^e) = 0$. The use of flow densities is advised for networks with higher dimensions. In the study of self-equilibrated networks, special interest is given to flow variables and the flow space $\mathcal{F}$, with space $\mathcal{F}$ referring to the actual flow space in one-dimensional applications and to the flow density space in higher dimensions.

## 3. Self-equilibrated network models

In network equilibrium models, cut laws and circuit laws are always defined by linear systems as attested by Equations 7 and 10. The cut laws and the circuit laws can thus be expressed as:

$$A^{\delta p} \cdot \omega^f = F \qquad (11)$$

$$A^f \cdot \omega^{\delta p} = P \qquad (12)$$

where the effects of the external perturbations to the system are lumped into the vectors $F$ and $P$. The solutions to Equations 11 and 12 admit two parts, a homogeneous solution that depends solely on the topology of the system and the values assigned to the other variable type, and a particular solution that depends also on the external perturbation. Algebraically, the homogeneous solution for the flow density $\omega^f$ and the potential density $\omega^{\delta p}$ corresponds to the nullspace of the matrices $A^{\delta p}$ and $A^f$. Self-equilibrium occurs when the system is in a non-trivial equilibrium state in the absence of external perturbations ($F = 0$ and $P = 0$).

A self-equilibrated network is thus a network where the cut-laws have a non-zero homogeneous solution reflecting that the system can be in a state of self-equilibrium in the absence of external perturbations. Let $\Omega^f$ be the collection of the $s$ basis vectors of the null-space of $A^{\delta p}$ and $\alpha \in \mathbb{R}^s$ be a vector of $s$ real coefficients. The flow density solution can be expressed as:

$$\omega^f = \omega_p^f + \omega_h^f = \omega_p^f + \Omega^f \alpha \tag{13}$$

In this case, cut laws admit an infinity of solutions governed by the nullspace of the flow equilibrium matrix. This is important when studying the redundancy of the network and its ability to sustain damage. The flow equilibrium matrix $A^{\delta p}$ is a $dn_v \times n_e$ matrix where each column corresponds to an edge and each row describes an elementary cut of a node in a given dimension. The nullspace of $A^{\delta p}$ exists if, and only if, $A^{\delta p}$ has redundant columns (edges). Therefore, the existence of a flow mode in the network reflects that the network has more edges than required for flow admission. In other words, each vector in $\Omega^f$ corresponds to a different flow path inside the network. The different flow modes correspond to different independent cycles of the network graph when the potential is one-dimensional. For electrical circuits, this indicates the existence of duplicate components and that the loss of one component does not necessarily lead to the failure of the electrical circuit as a whole. In structural systems, the existence of multiple flow modes indicates that the structure is indeterminate having multiple load paths, while in transportation networks, multiple flow modes reflect the existence of multiple distribution paths from one point to another.

Analogously, the potential equilibrium matrix $A^f$ can have a nullspace depending on its rank. In this case, for a given topology and flow vector, the circuit laws admit an infinite number of solutions governed by the nullspace of the potential equilibrium matrix $A^f$. Let $\Omega^{\delta p}$ be the collection of the $t$ basis vectors of $A^f$, then:

$$\omega^{\delta p} = \omega_p^{\delta p} + \omega_h^{\delta p} = \omega_p^{\delta p} + \Omega^{\delta p} \alpha \tag{14}$$

The existence of these infinite solutions reflects that a given flow could be associated to multiple potentials, hence the potential can continuously change without affecting the flow. An example of this feature can be found in the structural analysis of self-stressed networks, where the existence of potential modes may indicate the existence of infinitesimal mechanism in the structure.

### 3.1. Cellular structure of self-equilibrated networks

As established in previous sections, the algebraic structure of the cut laws and circuit laws solutions forms a vector space. Consequently, the behavior of the network equilibrium model subject to external perturbations and/or damage (i.e., member removal) can be predicted by analyzing the topology of the network and the potential at each node. De Guzmán and Orden mathematically proved that all self-equilibrated frameworks are composed of elementary cells [38], while Aloui et al. developed a bio-inspired generative approach to design and analyze self-stressed frameworks embedded in two-dimensional and three-dimensional spaces by decomposing the underlying graph to elementary units called cells [35-37]. They showed that the basis for the flow space $\mathcal{F}$ can be described solely by getting the cellular structure of the graph. In this paper, the approach is generalized to any arbitrary potential dimension.

Let $G(V, E)$ be a graph that describes the set of nodes $V$ and the set of edges $E$ of a system. Consider $f$ as the flow variable attributed to the edges, $\omega$ as the corresponding flow density, and $p$ as the potential attributed to the nodes. The graph $G$ is embedded in a $d$-dimensional space. A cell is defined as the complete graph on $d + 2$ nodes that has a one-dimensional flow space. Figure 2 shows the cell topology in a one-, two-, three- and four-dimensional space.

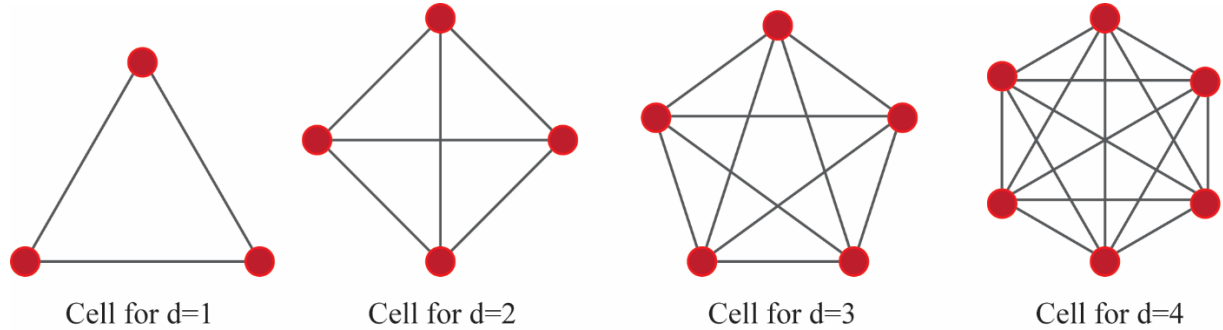

*Figure 2: Illustration of cell topologies in a one-, two-, three- and four-dimensional space.*

Complex self-equilibrated frameworks can be obtained through cellular multiplication and the mechanisms of adhesion and fusion (Figure 3) [35-37]. Adhesion represents the combination of two cells without removing any shared edges (the underlying graphs are glued together), while fusion refers to the combination of two cells with the removal of one or more of their shared edges. Consequently, adhesion increases the flow space dimension $\mathcal{F}$ and the number of flow modes in the network, while fusion reduces the dimension of the flow space $\mathcal{F}$ of the network and the number of flow modes. It should also be noted that in cellular morphogenesis, there is a distinction between cells and unicellular organisms, with cells having always the same topological structure (a complete graph), while unicellular organisms represent network structures with one flow mode. Unicellular organisms are required to obtain a complete description of the flow space.

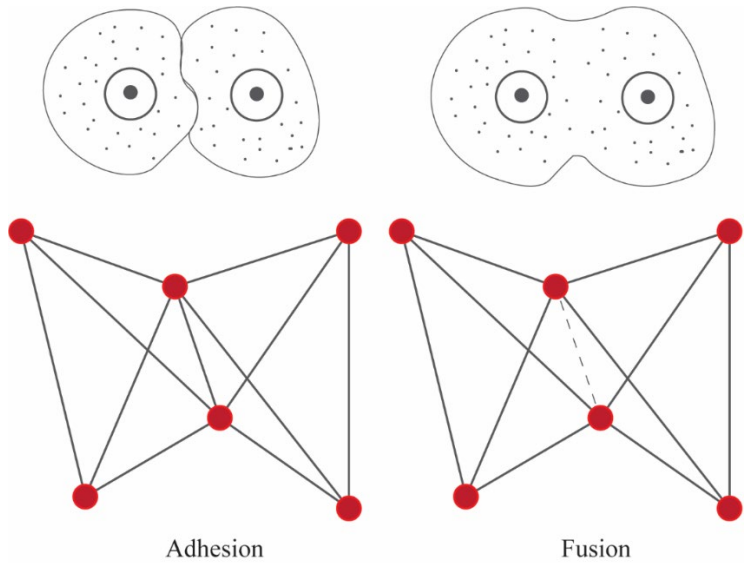

*Figure 3: Illustration of the adhesion and fusion mechanisms using two-dimensional cells.*

The cellular structure of a self-equilibrated network model refers to the series of cells and unicellular organisms that through adhesion compose the network. A decomposition algorithm for self-equilibrated network models, that gives a cellular structure of a $d$-dimensional network model based on $d$-dimensional cells was proposed in [37]. Figure 4 shows the cellular structure of two examples of network models that have the same underlying graph but are embedded in two different dimensions ($d = 1, d = 2$).

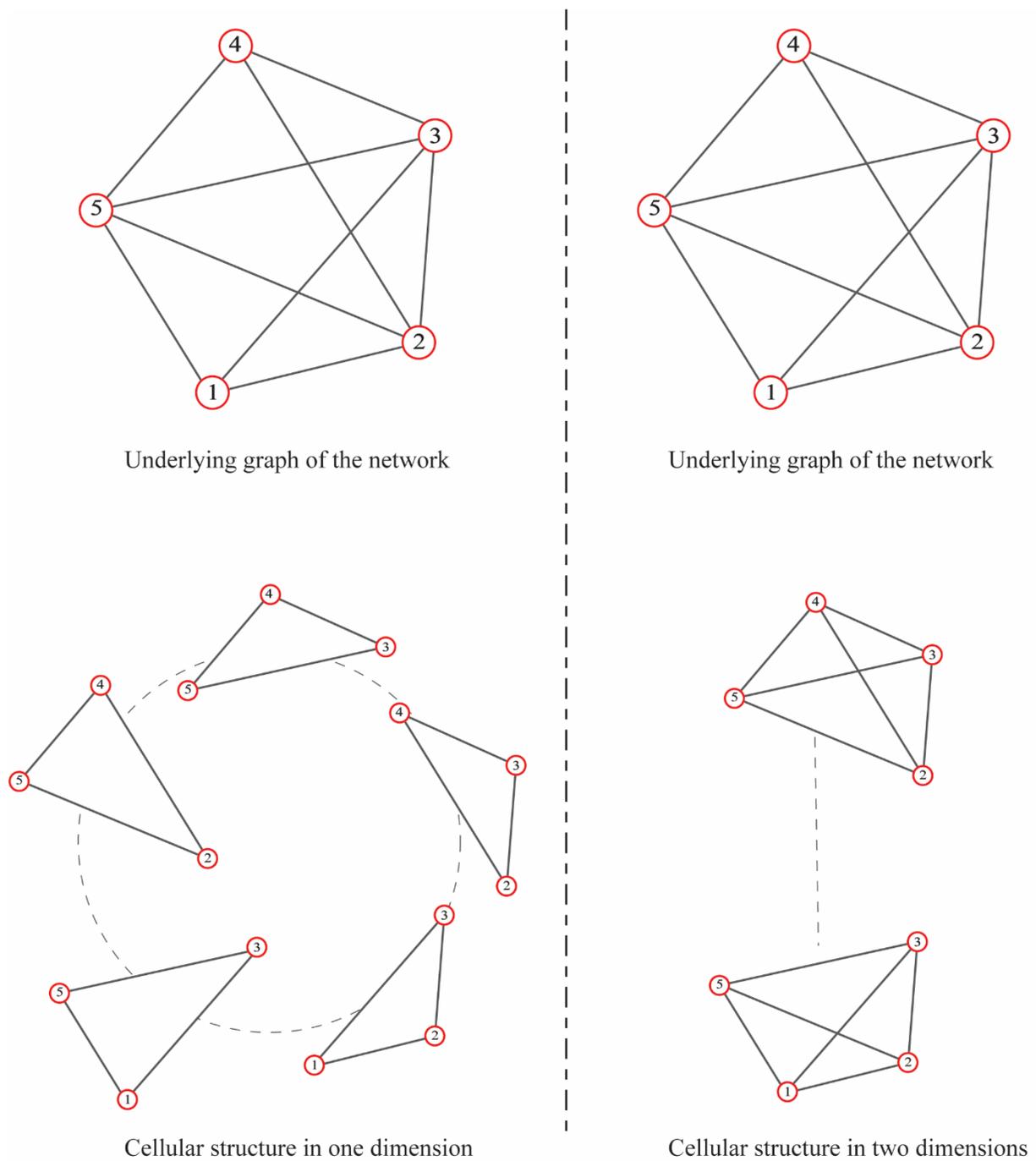

*Figure 4: Cellular structure of a one-dimensional embedding (left) and two-dimensional embedding (right) of the same network.*

3.2. Flow modes and space

Flow modes are defined as the vector basis for the flow space $\mathcal{F}$ of a self-equilibrated network. They thus represent the vector solution for the flow variables that satisfy the cut laws. In cellular morphogenesis, flow modes have a more direct interpretation with every flow mode corresponding to a cell or unicellular organism composing the network. In previous work, Aloui et al. proposed an analytical solution for the flow mode of a two-dimensional and a three-dimensional cell [33, 34]. They showed that the flow densities of the cells in two dimensions and three dimensions can be obtained through the product of the signed volume of two specific oriented two-simplices and three-simplices, respectively. In this paper, using algebraic geometry, a generalization of the analytical solutions for the flow mode for cells of an arbitrary dimension is proposed. A detailed proof of the generalization can be found in Appendix B. It is shown that, in a $d$-dimensional space, this result can be generalized to the product of the signed volumes of two specific oriented $d$-simplices each defined by an ordered set of $d + 1$ nodes in the cell. Let $\omega_{ij}$ be the flow density of the edge $(v_i, v_j)$, $S_i = \{v_i, (v_k)_{\substack{1 \leq k \leq d+2 \\ k \neq i \neq j}}\}$ and $S_j = \{v_j, (v_k)_{\substack{1 \leq k \leq d+2 \\ k \neq i \neq j}}\}$ the two ordered sets of vertices representing the $d$-simplices at $v_i$ and $v_j$, and $V(S)$ the function that returns the oriented volume of the oriented simplex $S = \{v_{\delta_1}, \ldots v_{\delta_{d+1}}\}$ where $\{\delta_1, \delta_2, \ldots, \delta_{d+1}\}$ is a specific order of the nodes. $P_k = (P_{1\delta}, \ldots, P_{d\delta})$ is the d-dimensional potential associated to $v_\delta$. The signed volume $V(S)$ is thus given by:

$$V(S) = \frac{1}{d!} \begin{vmatrix} 1 & P_{1\delta_1} & \cdots & P_{d\delta_1} \\ 1 & P_{1\delta_2} & \cdots & P_{d\delta_2} \\ \vdots & \vdots & \vdots & \vdots \\ 1 & P_{1\delta_{d+1}} & \cdots & P_{d\delta_{d+1}} \end{vmatrix} \quad (15)$$

Consequently, the flow density $\omega_{ij}$ can be obtained as:

$$\omega_{ij} = V\left(v_i, (v_\delta)_{\substack{1 \leq \delta \leq d+2 \\ \delta \neq i \neq j}}\right) V\left(v_j, (v_\delta)_{\substack{1 \leq \delta \leq d+2 \\ \delta \neq i \neq j}}\right) \quad (16)$$

By applying Equation 16 on all edges $(v_i, v_j) \in E$, one can obtain the analytical expression of the flow mode for a cell in a $d$-dimensional space.

Once an analytical solution for the flow density modes of the cells is obtained, a basis for the flow space $\mathcal{F}$ can be constructed considering the cellular structure of the network model. Each cell composing the network has its own flow mode and represents a component of the basis of the flow space $\mathcal{F}$. However, for the basis to be complete flow modes corresponding to unicellular structures have to also be considered. Flow modes corresponding to unicellular structures can be calculated using the fusion principles. When two cells undergo fusion (removal of one or more shared edges), the resulting network will have one flow density mode. Since each cell has one flow density mode, fusion can be thought of as finding the specific linear combination of flow modes of the two cells that cancels the flow density in the removed edges. Since every flow mode is defined to a constant, finding this specific combination is always possible when a single edge needs to be removed. However, when the number of removed edges is larger or equal to two, the potentials attributed to the nodes become degenerate collapsing to a lower dimensional space. Figures 7 and 8 show the cellular structures and flow density spaces $\mathcal{F}$ of the examples presented in the previous section (Figure 4) with flow density modes grouped into matrix $\Omega$.

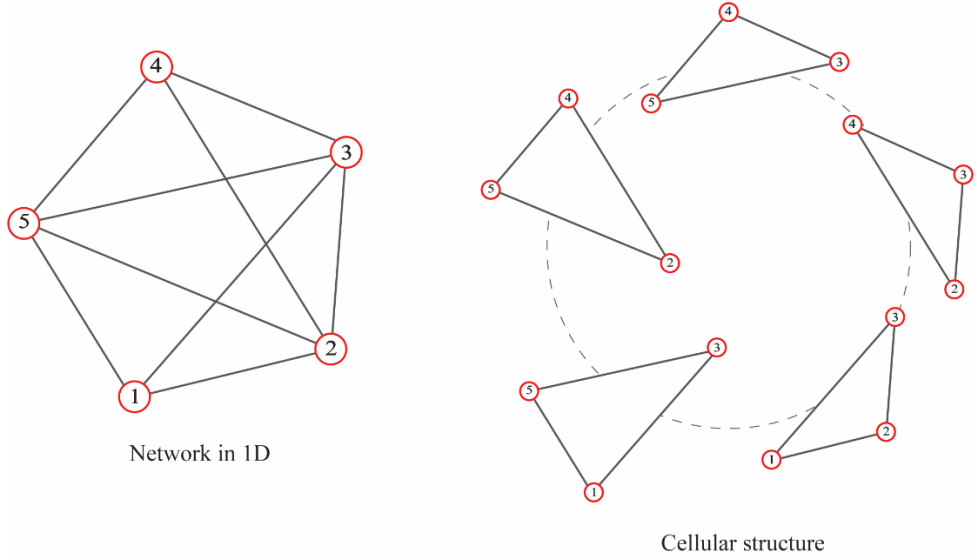

$$\Omega = \begin{bmatrix} \overbrace{V(v_1,v_3)V(v_2,v_3)}^{Cell\ \{v_1,v_2,v_3\}} & \overbrace{0}^{Cell\ \{v_1,v_3,v_5\}} & \overbrace{0}^{Cell\ \{v_2,v_3,v_4\}} & \overbrace{0}^{Cell\ \{v_2,v_4,v_5\}} & \overbrace{0}^{Cell\ \{v_3,v_4,v_5\}} \\ V(v_1,v_2)V(v_3,v_2) & V(v_1,v_5)V(v_3,v_5) & 0 & 0 & 0 \\ 0 & V(v_1,v_3)V(v_5,v_3) & 0 & 0 & 0 \\ V(v_2,v_1)V(v_3,v_1) & 0 & V(v_2,v_4)V(v_3,v_4) & 0 & 0 \\ 0 & 0 & V(v_2,v_3)V(v_4,v_3) & V(v_2,v_5)V(v_4,v_5) & 0 \\ 0 & 0 & 0 & V(v_2,v_4)V(v_5,v_4) & 0 \\ 0 & 0 & V(v_3,v_2)V(v_4,v_2) & 0 & V(v_3,v_5)V(v_4,v_5) \\ 0 & V(v_3,v_1)V(v_5,v_1) & 0 & 0 & V(v_3,v_4)V(v_5,v_4) \\ 0 & 0 & 0 & V(v_4,v_2)V(v_5,v_2) & V(v_4,v_3)V(v_5,v_3) \end{bmatrix} \begin{matrix} (v_1,v_2) \\ (v_1,v_3) \\ (v_1,v_5) \\ (v_2,v_3) \\ (v_2,v_4) \\ (v_2,v_5) \\ (v_3,v_4) \\ (v_3,v_5) \\ (v_4,v_5) \end{matrix}$$

Flow density modes and corresponding cells

*Figure 5: Cellular structure and flow density space $\mathcal{F}$ of a network embedded in a one-dimensional space.*

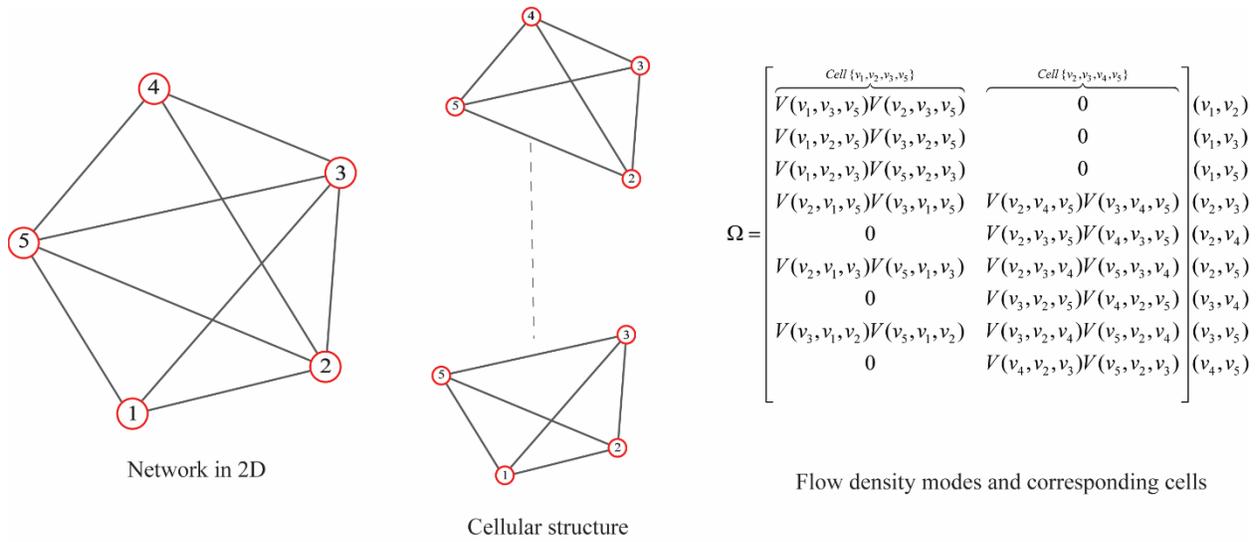

*Figure 6: Cellular structure and flow density space ℱ of a network embedded in a two-dimensional space.*

## 4. Equilibrium and damage analysis in self-equilibrated networks

In this section, equilibrium and damage in self-equilibrated networks are discussed through a series of examples. The examples, selected for their generality and brevity, represent network applications from different fields to highlight the applicability of the method in different contexts.

### 4.1. Equilibrium of self-equilibrated networks in the absence of external perturbation

Consider the electrical circuit illustrated in Figure 7. This circuit is known as the Wheatstone bridge and it is frequently used in electrical engineering for the precise measurement of an unknown electrical resistance through the balancing of two "arms" of a bridge circuit. It is also often used along with an operational amplifier to measure physical parameters such as temperature or strain, while variations of the bridge can also measure capacitance, inductance and impedance. Here, the topology of the circuit and its equilibrium state are studied through the analysis of the cellular structure of the network corresponding to the circuit (Figure 7).

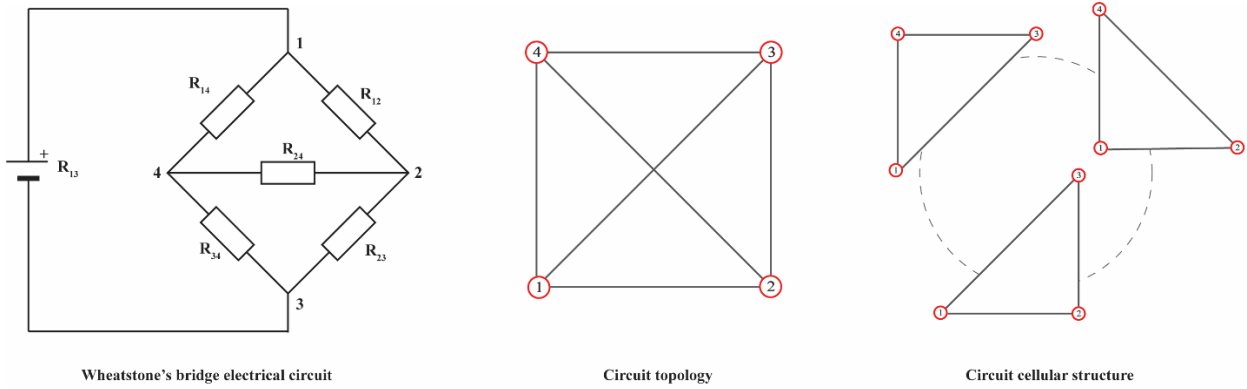

Figure 7: Wheatstone bridge electrical circuit and its topology.

The network equilibrium model corresponding to Wheatstone bridge is one-dimensional, with each node $v_i$ being associated with a scalar voltage potential $V_i$. Moreover, flow attributes in this model corresponds to the currents $I_{ij}$ in each edge $(v_i, v_j)$. Figure 8 shows the cells composing the network and the corresponding flow density modes calculated through the expressions developed in Section 3.2.

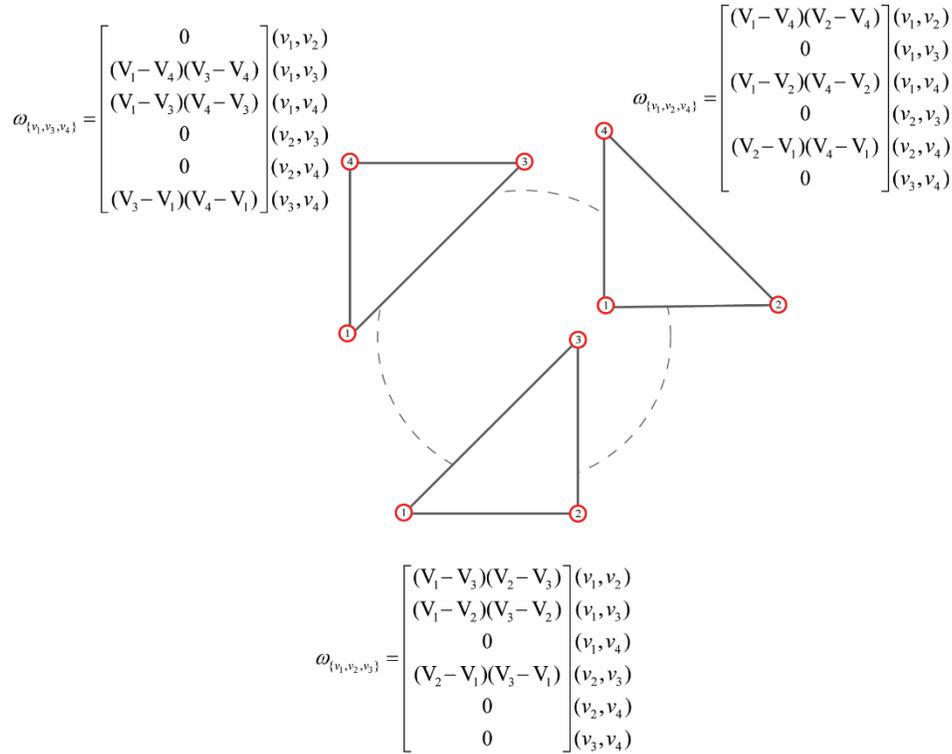

Figure 8: Cellular structure and flow density modes for the network equilibrium model corresponding to Wheatstone bridge electrical circuit.

The dimension of the flow density space $\mathcal{F}$ is three, given by the number of cells and unicellular structures composing the network. Any current density that can flow in the circuit is thus a linear combination of the three flow density modes given in Figure 8. This implies that the circuit is redundant and can withstand the loss of up to two edges provided that the damage does not affect the current or the voltage source. With each edge $(v_i, v_j)$ characterized by its own impedance $R_{ij}$, the network equilibrium model of Wheatstone bridge is described by:

| |   |
|---|---|
| Cut-set laws (current laws):<br>$$\sum_{j \text{ s.t. } (i,j) \in E} I_{ij} = 0, 1 \leq i \leq 4$$ | (17) |
| Circuit laws (voltage laws):<br>$$\sum_{(i,j) \in C} (V_i - V_j) = 0, C \in [[1,2,3], [1,2,4], [1,3,4]]$$ | |
| Flow and potential relation (Ohm's law):<br>$$r(\delta V, I) = \delta V - RI = 0$$ | |

$E$ represents the set of edges of the circuit network. $C$ refers to a cycle in the cycle space of the network. $R$ is a diagonal $6 \times 6$ matrix where each diagonal entry is the impedance $R_{ij}$ of the corresponding edge $(v_i, v_j)$. In a matrix form, the self-equilibrium of the circuit is described as:

| |   |
|---|---|
| Cut-set laws (current laws):<br>$$BI = \begin{pmatrix} 1 & 1 & 1 & 0 & 0 & 0 \\ 1 & 0 & 0 & -1 & -1 & 0 \\ 0 & -1 & 0 & -1 & 0 & 1 \end{pmatrix} \begin{pmatrix} I_{12} \\ I_{13} \\ I_{14} \\ I_{23} \\ I_{24} \\ I_{34} \end{pmatrix} = 0$$ | (18) |
| Circuit laws (voltage laws):<br>$$CI = \begin{pmatrix} 1 & -1 & 0 & 1 & 0 & 0 \\ 1 & 0 & -1 & 0 & 1 & 0 \\ 0 & 1 & -1 & 0 & 0 & 1 \end{pmatrix} \begin{pmatrix} V_1 - V_2 \\ V_1 - V_3 \\ V_1 - V_4 \\ V_2 - V_3 \\ V_2 - V_4 \\ V_3 - V_4 \end{pmatrix} = 0$$ | |
| Flow and potential relation (Ohm's law):<br>$$r(\delta V, I) = \begin{pmatrix} V_1 - V_2 \\ V_1 - V_3 \\ V_1 - V_4 \\ V_2 - V_3 \\ V_2 - V_4 \\ V_3 - V_4 \end{pmatrix} - \begin{pmatrix} R_{12} & 0 & 0 & 0 & 0 & 0 \\ 0 & R_{13} & 0 & 0 & 0 & 0 \\ 0 & 0 & R_{14} & 0 & 0 & 0 \\ 0 & 0 & 0 & R_{23} & 0 & 0 \\ 0 & 0 & 0 & 0 & R_{24} & 0 \\ 0 & 0 & 0 & 0 & 0 & R_{34} \end{pmatrix} \begin{pmatrix} I_{12} \\ I_{13} \\ I_{14} \\ I_{23} \\ I_{24} \\ I_{34} \end{pmatrix} = 0$$ | |

The self-equilibrium of the circuit in Equation 18 can thus be used to explain the principle of the Wheatstone bridge which is based on null deflection.

### 4.2. Equilibrium of self-equilibrated networks under external perturbation

The example of a self-stressed three-dimensional pin-jointed framework consisting of elements in compression (bars) and elements in tension (cables) is analyzed using network equilibrium modeling. Self-equilibrated axially loaded structures, also known as tensegrity structures, have been proposed for a variety of applications in science and engineering from cellular modeling to robotics. Moreover, they are statically indeterminate structures (i.e. they contain multiple load paths) with often multiple self-stress states. Here, the proposed framework consists of two three-dimensional cells combined through adhesion. For simplicity, all bars and cables are assumed to have the same cylindrical cross-sectional areas of 10 cm² and 1 cm², respectively. They also have the same material properties with a Young modulus $E_{bar}$=210 $GPa$ and $E_{cable} = 30 GPa$. Figure 9 illustrates the configuration and the type of elements in the structure while Figure 10 shows its corresponding underlying graph along with the three-dimensional cellular structure.

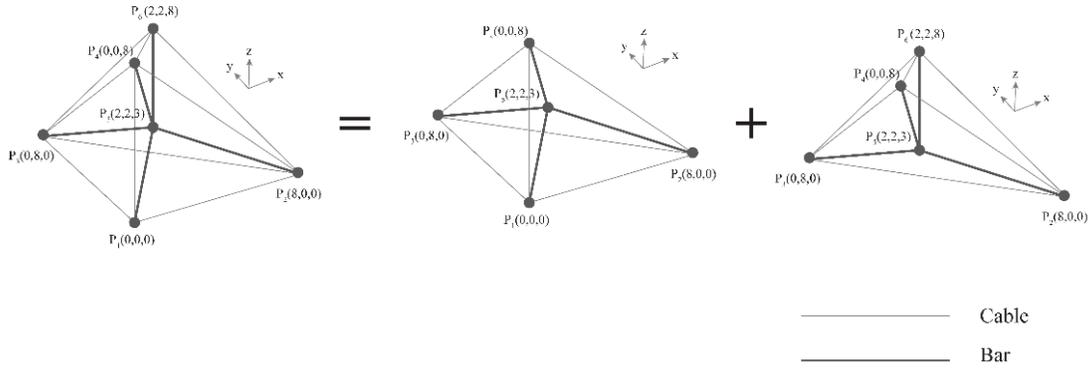

Figure 9: Configuration of the structure embedded in a three-dimensional space along with its cellular structure.

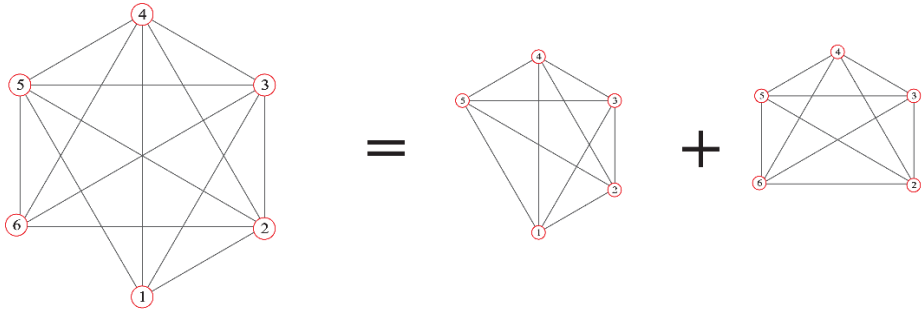

Figure 10: Underlying abstract graphs for the structure illustrated in Figure 9.

The cellular structure of this three-dimensional network model reveals that the structure is composed of two cells. Consequently, the flow density space has two flow density modes. The cells and their corresponding flow density modes are shown in Figure 11. The flow density variable in this case corresponds to the self-stress inside the structure. Any self-stress state of the structure will thus be a linear combination of these two self-stress modes. The model can thus explain and decipher statical indeterminacy in a tensegrity structure.

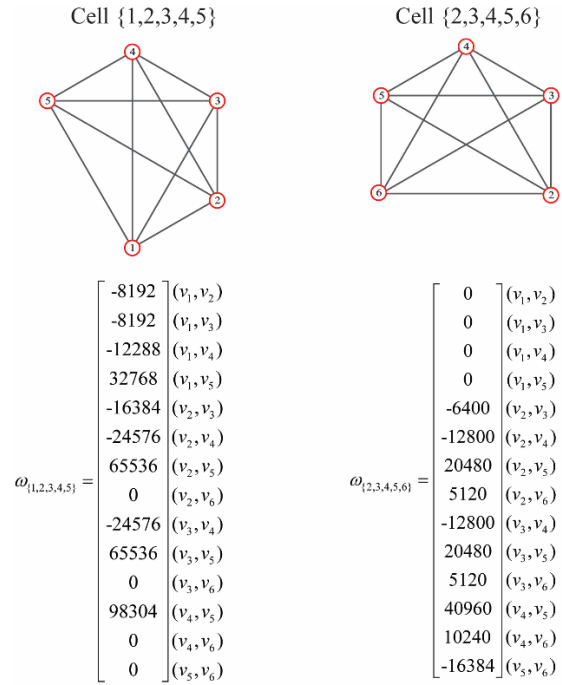

$$\omega_{\{1,2,3,4,5\}} = \begin{bmatrix} -8192 \\ -8192 \\ -12288 \\ 32768 \\ -16384 \\ -24576 \\ 65536 \\ 0 \\ -24576 \\ 65536 \\ 0 \\ 98304 \\ 0 \\ 0 \end{bmatrix} \begin{matrix} (v_1,v_2) \\ (v_1,v_3) \\ (v_1,v_4) \\ (v_1,v_5) \\ (v_2,v_3) \\ (v_2,v_4) \\ (v_2,v_5) \\ (v_2,v_6) \\ (v_3,v_4) \\ (v_3,v_5) \\ (v_3,v_6) \\ (v_4,v_5) \\ (v_4,v_6) \\ (v_5,v_6) \end{matrix} \qquad \omega_{\{2,3,4,5,6\}} = \begin{bmatrix} 0 \\ 0 \\ 0 \\ 0 \\ -6400 \\ -12800 \\ 20480 \\ 5120 \\ -12800 \\ 20480 \\ 5120 \\ 40960 \\ 10240 \\ -16384 \end{bmatrix} \begin{matrix} (v_1,v_2) \\ (v_1,v_3) \\ (v_1,v_4) \\ (v_1,v_5) \\ (v_2,v_3) \\ (v_2,v_4) \\ (v_2,v_5) \\ (v_2,v_6) \\ (v_3,v_4) \\ (v_3,v_5) \\ (v_3,v_6) \\ (v_4,v_5) \\ (v_4,v_6) \\ (v_5,v_6) \end{matrix}$$

*Figure 11: Flow density modes of the structure in the initial configuration*

The structure is initially in equilibrium under the effect of prestress introduced by applying a relative shortening $\frac{\Delta l}{l_0}$ of the cables of $10^{-3}$ (Figure 12). This state of self-equilibrium is described by the vector of internal forces $f$ which can be expressed as a linear combination of the flow modes.

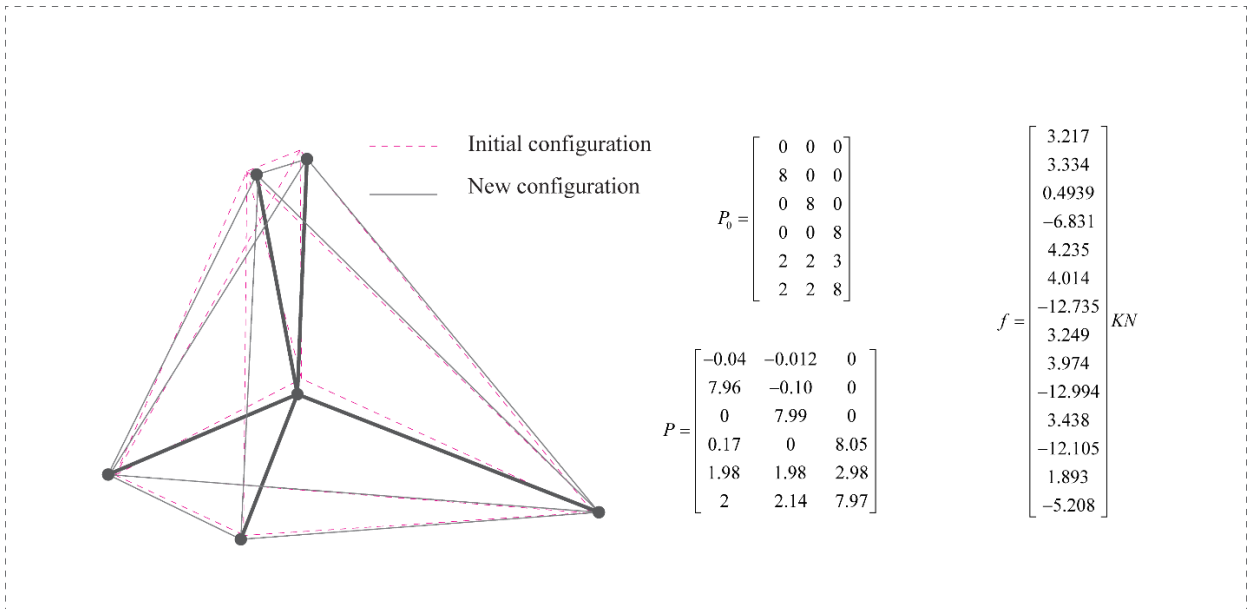

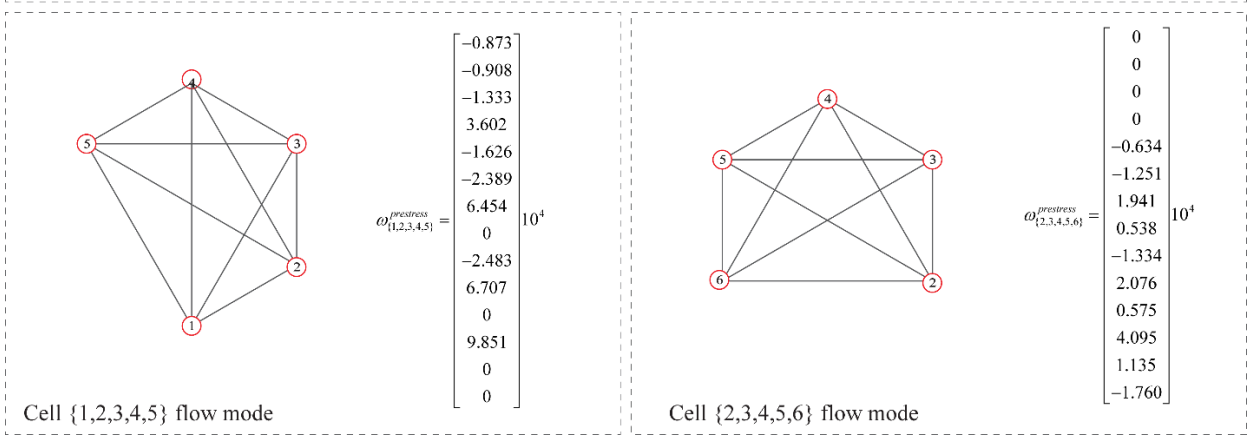

*Figure 12: Illustration of the new equilibrium configuration under prestress and the corresponding flow modes*

Now, consider that an external load $\vec{F} = -10\,\vec{z}\,KN$ is applied to the structure at node $v_6$ with the vertical displacements of nodes $v_1$, $v_2$ and $v_3$ blocked generating the reactions $\vec{R}_1 = 4.8\,\vec{z}\,KN$, $\vec{R}_2 = 2.6\,\vec{z}\,KN$ and $\vec{R}_3 = 2.6\,\vec{z}\,KN$ (Figure 13). The equilibrium of the structure implies that the sum of applied forces and reactions is equal to zero and that the sum of the moments with respect to a point in space is zero. The network equilibrium model for this system is given by the nodal equilibrium and geometric compatibility equations representing cut-set and circuit laws, respectively, while Hooke's laws applied at each element of the structure correspond to the potential-flow relations (Equation 19).

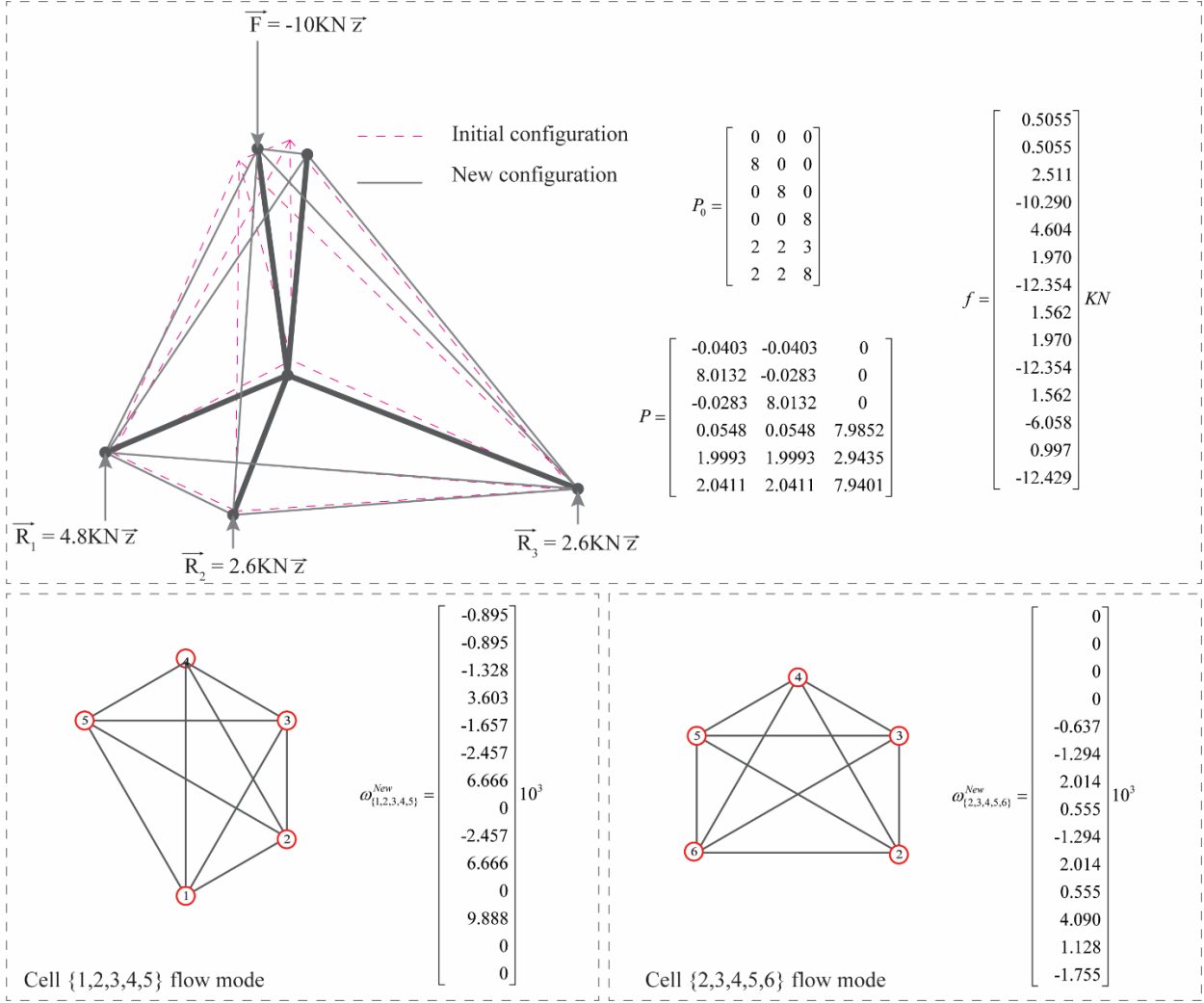

*Figure 13: Illustration of the new equilibrium configuration under perturbation and the corresponding flow modes.*

Cut laws can be determined by isolating the nodes through elementary cut-sets. Circuit laws have to be expressed in a basis of the cycle space $\mathcal{C}$. Since cycles can be viewed as one-dimensional cells and unicellular organisms, a basis for the cycle space $\mathcal{C}$ can be found through the cellular structure of the corresponding graph. Figure 14 illustrates the cycle space basis for the example.

| Cut-set laws (nodal equilibrium): | |
|---|---|
| $$\sum_{j \text{ s.t.} (i,j)\in E} \vec{f}_{ij} = F_j, \quad 1 \leq i \leq 6$$ | |
| Circuit laws (geometric compatibility): | |
| $$\sum_{(i,j)\in \mathcal{C}} (\delta \vec{P}_i - \delta \vec{P}_j) = 0, \quad \mathcal{C} \in [\{1,2,3\},\{1,3,4\},\{1,4,5\},\{2,3,4\},\{2,4,5\},\{2,5,6\},\{3,4,5\},\{3,5,6\},\{4,5,6\}]$$ | (19) |
| Flow and potential relation (Hooke's law): | |
| $$f_{(i,j)} - K_{(i,j)} \underbrace{(\|\vec{P}_{(i,j)}\| - \|\vec{P}^0_{(i,j)}\|)}_{\delta l_{(i,j)} = l_{(i,j)} - l^0_{(i,j)}} = 0, \quad \forall\, (i,j) \in E$$ | |

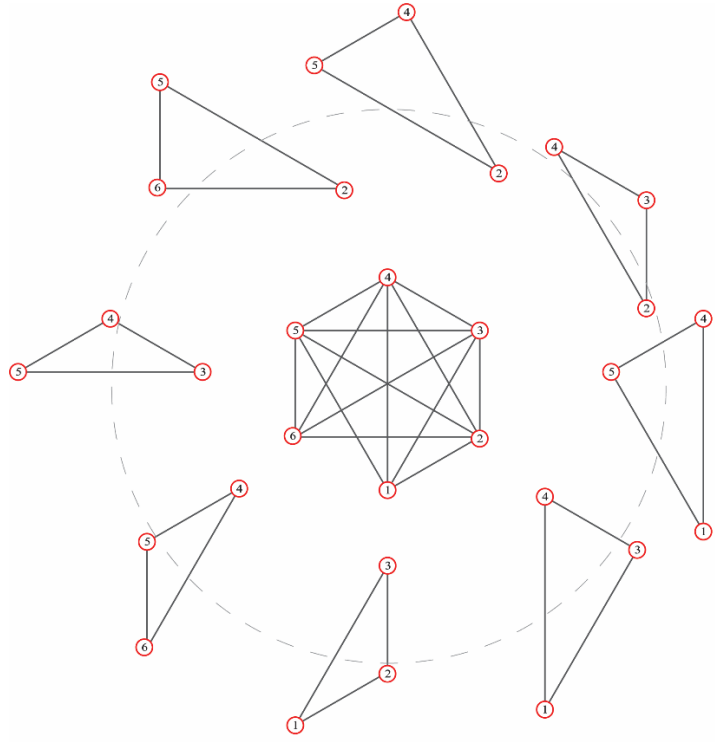

*Figure 14: Cycle-space basis based on one-dimensional cellular structure of the network.*

In Equation 19, $C$ represents a cycle basis of the cycle space $\mathcal{C}$. $\vec{P}_i$ is the position vector of node $v_i$. $\delta \vec{P}_i$ is the displacement of node $v_i$. $\vec{P}_{(i,j)}$ is the vector representation of the member $(i,j)$ subjected to perturbations. $f_{(i,j)}$ is the normal force of the member $(i,j)$. $\|\cdot\|$ is the Euclidean norm of a vector. $l_{(i,j)}$ and $l^0_{(i,j)}$ are the actual length and rest length of the member $(i,j)$ and $K_{(i,j)} = \dfrac{E_{(i,j)} A_{(i,j)}}{l_{(i,j)}}$ is the normal stiffness of member $(i,j)$ where $E_{(i,j)}$ is its Young modulus and $A_{(i,j)}$ is its cross section. In matrix form, the equilibrium of the structure is described by Equations 20.

| | |
|---|---|
| Cut-set laws (nodal equilibrium): $$A^{\delta p}\, \omega = F$$ $A^{\delta p}$ is the equilibrium matrix calculated using equation 15 | |
| Circuit laws (geometric compatibility): $$(\mathbb{I}_d \otimes C)\, \delta P = (\mathbb{I}_d \otimes C)\, [(\mathbb{I}_d \otimes B^T)\,(P - P^0)] = 0$$ $C$ is the cycle space matrix, $B$ is the node to branch adjacency matrix, $\delta P = \begin{bmatrix}\delta X \\ \delta Y \\ \delta Z\end{bmatrix}$ and $P^0 = \begin{bmatrix}X_0 \\ Y_0 \\ Z_0\end{bmatrix}$ where $\delta X, \delta Y$ and $\delta Z$ are the $6\times 1$ vectors of the nodal displacements and $X_0, Y_0$ and $Z_0$ are the $6\times 1$ vectors of the initial coordinates of the nodes. | (20) |
| Flow and potential relation (Hooke's law): $$f - K \underbrace{(\|P_{vec}\|_{vec} - \|P^0_{vec}\|_{vec})}_{\delta l = l - l_0} = 0$$ | |

> $f$ is a $14 \times 1$ vector of the internal forces of the members. $K$ is the stiffness matrix of the structure. $P_{vec} = [X_{vec}\ Y_{vec}\ Z_{vec}]$ and $P_{vec}^0 = [X_{vec}^0\ Y_{vec}^0\ Z_{vec}^0]$ where $X_{vec}, Y_{vec}$ and $Z_{vec}$ are $14 \times 1$ vectors reflecting the components of the vector representations of the members in the new configuration and $X_{vec}^0, Y_{vec}^0$ and $Z_{vec}^0$ are $14 \times 1$ vectors reflecting the components of the vector representations of the members in the initial configuration. $\|\cdot\|_{vec}$ is the row vector wise norm of a matrix. $l$ is a $14 \times 1$ vector representing the actual lengths of the members and $l_0$ represents the rest lengths of the members.

When the structural system is subjected to perturbations, the nodal positions will adjust to the new equilibrium. Consequently, the flow density space will change. However, the cellular structure of the network remains the same implying that the network has always two flow density modes. The equilibrium state of the network can thus be described through a homogeneous solution describing the self-equilibrium in the absence of external loading and a particular solution which reflects the effect of the load [37]. Since flow-potential relations $r(P, f)$ are non-linear with respect to the configuration (geometry) of the structure $P$, finding the equilibrium configuration of the structure under the effect of the external load requires the use of an appropriate numerical method [39-41]. In this paper, a dynamic relaxation algorithm [41] was employed to calculate the positions of the nodes and the internal forces in the elements of the structure under the effect of the external load $\vec{F}$ and the reactions $\vec{R}_1, \vec{R}_2$ and $\vec{R}_3$ considering all z-displacements on the basis nodes as blocked. In this analysis, self-weight was neglected and prestress in the cables was induced by elongation of 0.01 % of their rest length. Figure 13 illustrates the new equilibrium configuration and the new flow density modes. The model can thus be used to analyze self-stressable pin-jointed frameworks and explain their behavior under loading.

### 4.3. Equilibrium of self-equilibrated networks under damage

In this section, the impact of damage (element removal) is assessed in a self-equilibrated network through the analysis of its cellular structure. Element removal can be thought of as the result of a fusion on the removed edge. Consequently, the number of cells and thus the number of flow modes decrease, reducing also the redundancy in the network. Element removal can thus be reflected by the fusion or the necrosis of cell. In this example, a two-dimensional network is analyzed. Figure 15 describes the network, its associated potential, cellular structure and corresponding flow density modes.

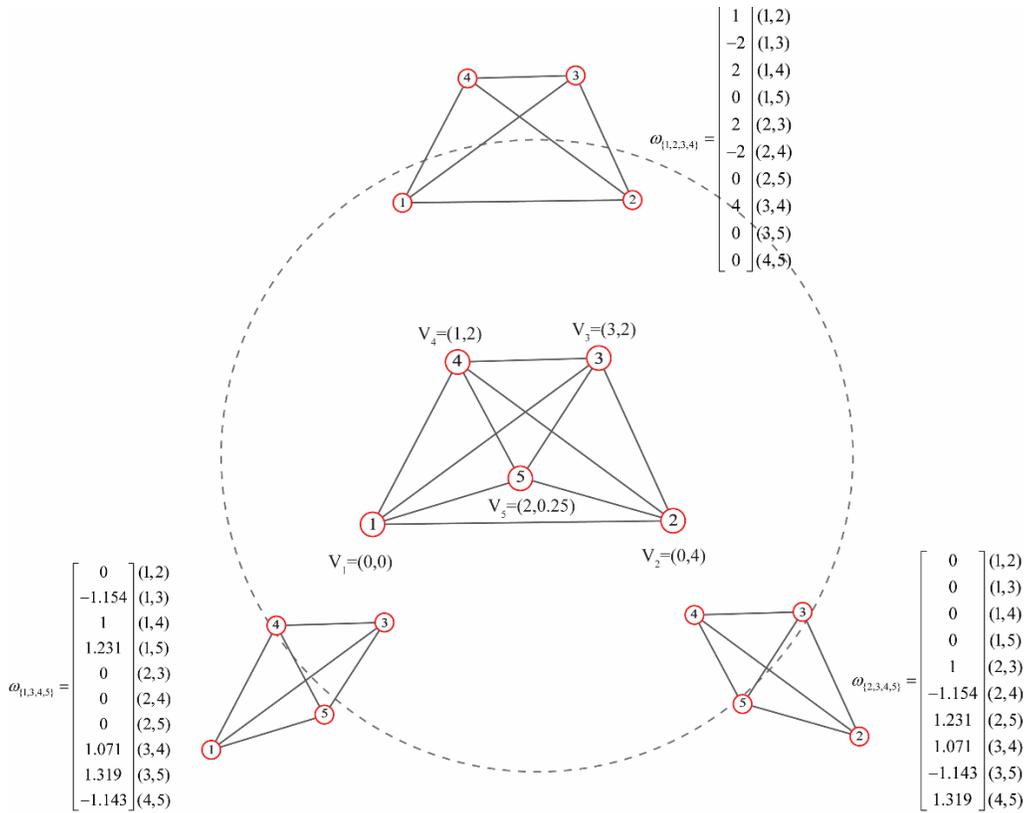

*Figure 15: Illustration of the underlying graph of the network along with its associated cellular structure and flow density modes.*

Let $\omega_0$ be the initial flow density that the network has under the potential values included in Figure 16. $\omega_0$ satisfies the self-equilibrium conditions and is a linear combination of the flow density modes provided in Figure 15: $\omega_0 = \omega_{\{1,2,3,4\}} + \omega_{\{1,3,4,5\}} + 2\omega_{\{2,3,4,5\}}$.

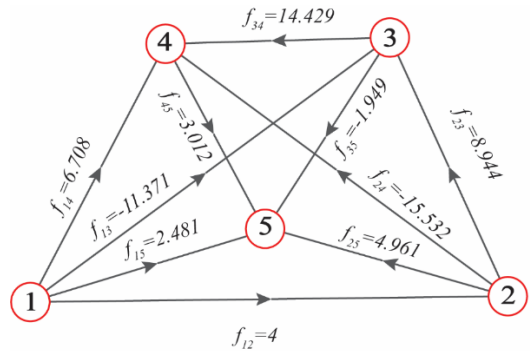

*Figure 16: Initial flow $f_0$ inside the network before damage.*

When an edge of the network is damaged (removed), the flow in the network can be adjusted according to where the damage occurs and to the system being modeled. Assume that edge (1,5) is removed. Topologically, the damage of edge (1,5) is described by the necrosis of cell {1,3,4,5}. The number of composing cells and thus the dimension of the flow space $\mathcal{F}$ will decrease from three to two. The same effect occurs with the damage of edge (3,4) which can be described topologically by the fusions of cells {1,2,3,4} + {1,3,4,5} and cells {1,2,3,4} + {2,3,4,5}. Figure 17 shows the cellular structure of the damaged networks, along with their corresponding flow density modes.

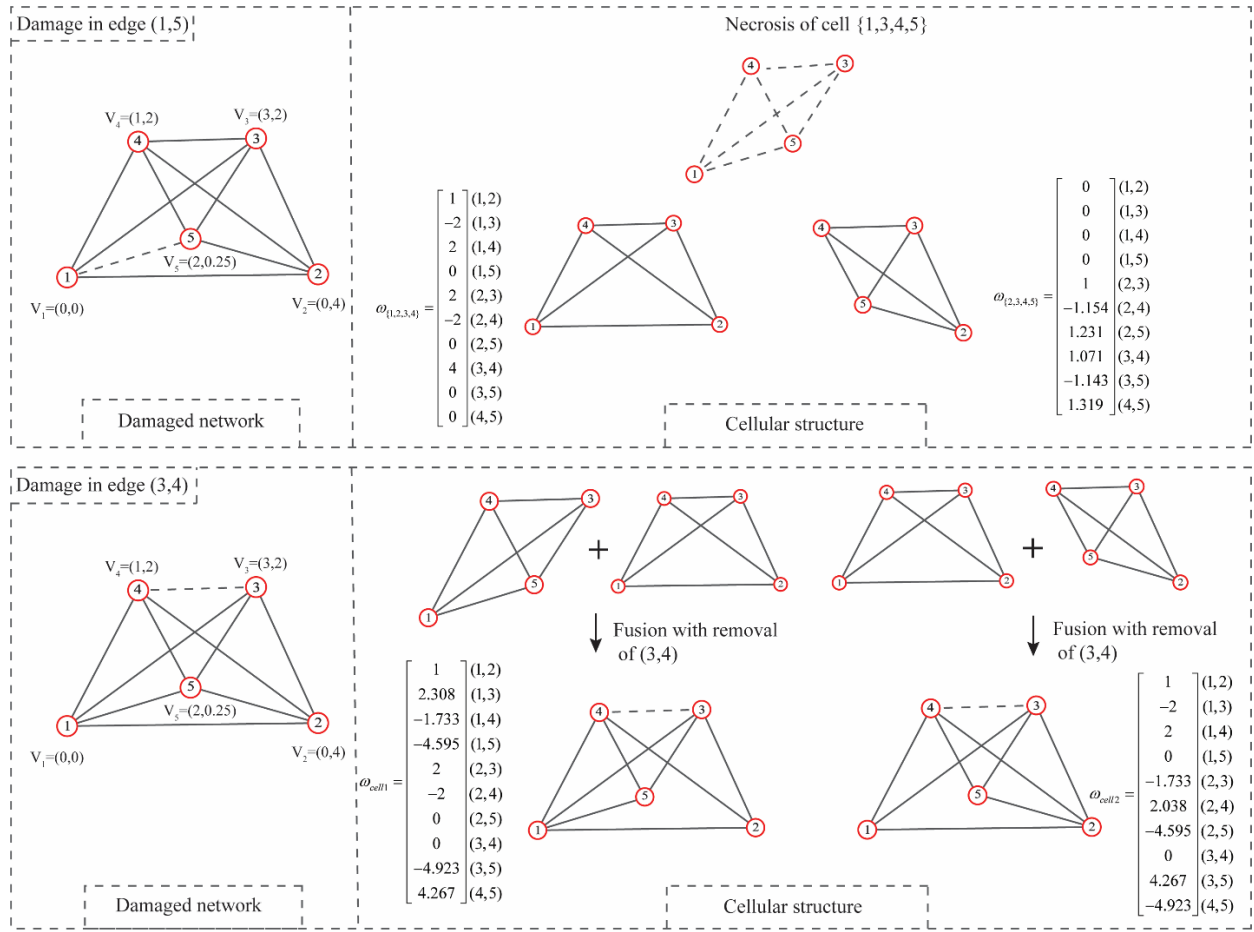

*Figure 17: Cellular structure and flow density modes in the case of damage in edges (1,5) and (3,4).*

Now, assume that the system being modeled has an additional constraint expressed by a desire to maintain flow at the same level before damage. Since the network has two redundant edges after damage, this is only possible for two edges. Considering the damage of edge (3,4), the flow in edges (1,5) and (2,5) can be kept the same with the new equilibrium given by:

$$\begin{pmatrix} -4.595 & 0 \\ 0 & -4.595 \end{pmatrix} \begin{pmatrix} \alpha_1 \\ \alpha_2 \end{pmatrix} = \begin{pmatrix} \frac{2.481}{2.016} \\ \frac{4.961}{2.016} \end{pmatrix} \qquad (21)$$

which gives $\alpha_1 = -0.268$ and $\alpha_2 = -0.536$. Note that flow values are divided by the geometric length of the corresponding edge to get the flow densities. The new flow on the system is represented by Figure 18.

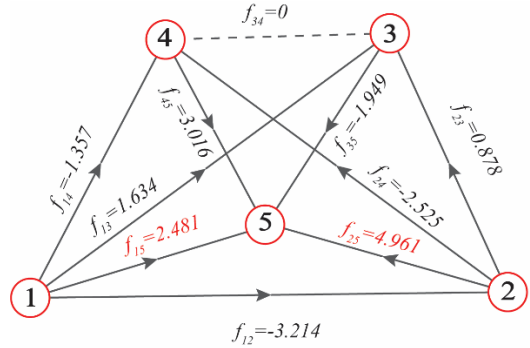

Figure 18: Flow after damage of edge (3,4) and with the consideration of maintaining flow in edges (1,5) and (2,5) at the same level prior to damage.

In this example, the network and the constraint are chosen for simplicity. However, in a real system the designer or decision maker can choose any constraint or objective function to direct flow distribution with the flow problem, after damage being reduced into the identification of the appropriate $\alpha_1$ and $\alpha_2$ instead of optimizing with respect to the nine flow variables at each edge. Moreover, it should also be noted that there are systems, like the pin-jointed framework analyzed in the previous example, where the potential and flow values are closely related, adjusting simultaneously to a new equilibrium position after damage. In these cases, flow-potential relations can be used to simulate change and find the new equilibrium configuration of the network. However, changes in the number of flow density modes can already be identified through a review of the cellular structure of the network.

## 5. Discussion

In summary, the paper focuses on self-equilibrated networks offering a new approach for their modeling and analysis. The proposed model is composed by a set of cut-set and circuit laws along with a set of flow and potential relations, while its analysis is conducted in terms of the cellular structure of the network. Cells refer to unitary network sub-systems that have a one-dimensional flow space with their dimension depending on the embedding of the network. An analytical expression of the flow mode of cells in a $d$-dimensional space is provided. This expression combined with the cellular structure of a self-equilibrated network allows to analyze the equilibrium state in the network through the study of individual cell equilibria and their interactions. Self-equilibrated networks represent thus networks where cut-laws have a non-zero homogeneous solution. The system can thus be in a state of self-equilibrium in the absence of external perturbations with flow modes corresponding to different independent cycles of the network graph when the potential is one-dimensional. This implies damage-tolerance and resilience in the system as the network includes different flow paths.

Given the wide range of physical and engineering systems that are depicted through network models, this modeling approach can thus have great impact in a variety of contexts. Here, the model is explored to describe the equilibrium in three network applications selected for their generality and brevity. The first example is the well-known Wheatstone bridge which is employed to measure electrical resistance as well as other physical parameters. It is shown that the model can be used to explain the functioning principle of Wheatstone bridge. The second example corresponds to a self-stressable pin-jointed structure composed of bars and cables. This type of structures, also known as tensegrity, are statically indeterminate structures. It is shown that the model proposed can be used to explain their self-equilibrium as well as their behavior under loading. The third example focuses on the impact of damage (element removal) in a self-equilibrated network, where it is shown that the model can be used to identify changes in the number of flow density modes in the network through a simple review of its cellular structure. Since the cellular structure of the network is invariant of the load case and can accommodate the impact of damage (element removal) through cell fusion and/or necrosis, it provides an always topologically valid basis for the description for the network. Therefore, the proposed cellular approach represents a systematic and general approach for the analysis of self-equilibrated networks in science and engineering applications.

## 6. Acknowledgements and Funding Statement

This material is based upon work supported by the **National Science Foundation**, United States under Grant No. **1638336**. David Orden has been partially supported by project **MTM2017-83750-P** of the Spanish Ministry of Science (AEI/FEDER, UE), project **PID2019-104129GB-I00** of the Spanish Ministry of Science and Innovation, and by H2020-MSCA-RISE project, European Commission **734922 – CONNECT**, while Nizar Bel Hadj Ali gratefully acknowledges the financial support of the Fulbright Visiting Scholar Program for the academic year 2018-2019.

# Appendix A

Let $G(V, E)$ be a graph that describes the set of nodes $V$ and the set of edges $E$ of a network and $n_v$ be the number of vertices and $n_e$ the number of edges in the graph. The graph is equipped with a set of node and edge attributes that satisfy the equilibrium conditions referred to, in graph theory, as circuit laws and cut-set laws. In this paper, node attributes are referred to as potential attributes; they are denoted by a $n_v \times 1$ vector $p$ and they satisfy circuit laws. Note that each component $p_i$ represents the value of the potential at node $v_i$ which in turn is a $d \times 1$ vector where $d$ represents the dimensionality of the problem. Edge attributes are referred to as flow attributes; they are denoted by a $n_e \times 1$ vector $f$ and they satisfy cut-set laws.

## 7.1. Network laws

### 7.1.1. Cut-set laws

A cut-set is a partition of the vertices of the graph $V$ into two disjoint sets $S$ and $\bar{S}$. Cut sets can be described in the case of planar graphs by a closed cutting curve $\Delta$ that divides the plane of the graph into two regions: inside and outside (Figure 1). Analogously, If the graph is not planar, then $\Delta$ represents a closed hyper-surface dimension $d - 1$ if the graph is embedded in a $d$-dimnesional space. An orientation can be given to the cut $\Delta$: inside outside ($\Delta_1$ in Figure A.1) or the reverse ($\Delta_2$ in Figure A.1).

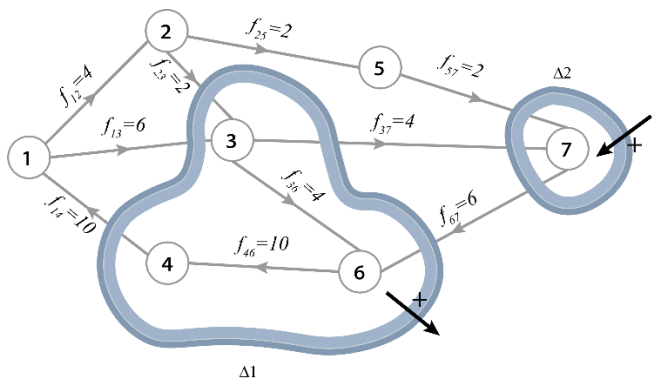

*Figure A.1: Illustration of a cut-set.*

Let $f: E \to \mathbb{R}^d$ be a real valued function on the edges of the graph $G$, where $d$ is the dimension of the problem. The values that $f$ takes on each edge $e \in E$ are referred to as the edge attributes of the graph. By convention, an edge attribute is positive if it is defined on an edge $e(u, v)$ with $u \in S$ and $v \in \bar{S}$, and negative if it is the opposite. In this paper, if $\Delta$ is a cut $[S, \bar{S}]$, $f(\Delta)$ will denote the sum of the edge attributes having one end node in $S$ and the other one in $\bar{S}$. Furthermore, $f$ obeys a cut-set law if, and only if, for any cut-set $\Delta = [S, \bar{S}]$, the edge attributes having one node in $S$ and the other node in $\bar{S}$ sum up to zero, i.e., $f(\Delta) = 0$.

### 7.1.2. Circuit laws

A circuit is a closed trail of the graph (i.e., a path with the first vertex being also the last one). A circuit $C$ is denoted by the set of nodes that forms it, taken in the order they are visited, and it can be represented by the set of edges in $C$ and an orientation (Figure A.2).

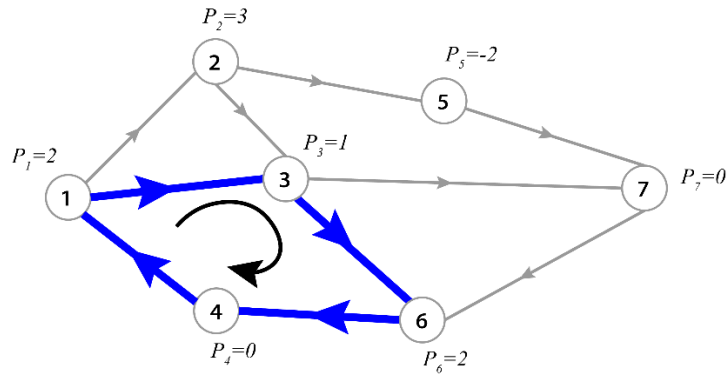

Figure A.2: Illustration of a circuit

Let $p: V \rightarrow \mathbb{R}^d$ be a real valued function on the vertices of the graph $G$, where $d$ is the dimension of the problem. The values that $p$ takes on each vertex are referred to as node attributes of the graph. Let $\delta p: E \rightarrow \mathbb{R}$ be the function that returns the node-attribute difference between the end nodes $u$ and $v$ of the edge $(u, v)$, i.e., $\delta p\big((u, v)\big) = p_u - p_v$. Such a $p$ obeys a circuit law if, and only if, for any circuit $C$ in the graph, the node attributes differences $\delta p$ sum up to zero over all the edges in the circuit, i.e., $\sum_{(u,v)\in C} \delta p\big((u, v)\big) = 0$.

7.2. Network attributes

In addition to its topological structure (nodal connectivity), a graph can be equipped with node and edge attributes that satisfy any given relations governed by the interconnections between the nodes (topology of the graph). When equilibrium is considered, these relations represent a network equilibrium model. Reinschke describes two types of attributes that occur in real-world systems that can be attributed to nodes and edges of a network equilibrium model [A.1] referred to in this paper as flow variables and potential variables.

7.2.1. Flow attributes $f$

A flow variable, by definition, is an edge attribute that satisfies a cut-set law on the graph. In Figure 1, $f$ is a flow variable. If $\Delta_1$ is considered, then $f\big((2,3)\big) + f\big((1,3)\big) + f\big((1,4)\big) + f\big((3,7)\big) + f\big((6,7)\big) = 0$. Examples of flow variables include currents in electrical systems, internal and external forces in structural systems, and fluxes of quantities in transportation problems. If the closed cut contains only one node, like $\Delta_2 = \{\{7\}, V\backslash\{7\}\}$ in Figure A.1, it is referred to as an elementary cut. In structural systems, the flow variable for a single node refers to the nodal equilibrium of forces, while in electrical circuits, it can be interpreted as Kirchhoff $1^{\text{st}}$ law (current law). In transportation, the cut law for a single node may describe the flow conservation.

### 7.2.2. Potential attributes $p$

A potential variable is a node attribute that satisfies a circuit law on the graph. In Figure A.2, $p$ is a potential. If circuit [1,2,6,4,1] is considered, then the circuit law describes the relation: $\delta p((1,3)) + \delta p((3,6)) + \delta p((6,4)) + \delta p((4,1)) = 0$. An elementary circuit for potential attributes represents a cycle. Circuit laws can be used to model Kirchhoff's second law (voltage law) in electrical circuits or geometric compatibility relations in structures. Consequently, potential variables can represent voltages in electrical circuits, nodal displacements in structural systems, or quantities of goods at a given node in a transportation scheme.

## 7.3. Network topological and algebraic properties

A network is described by a set of vertices $V$ and the connections $E$ between them. However, all information on the connectivity of the network can also be depicted using cut-sets or circuits. Let $G(V, E)$ be a directed graph where $V$ is the set of vertices, $n_v$ the number of nodes, $E$ is the set of edges and $n_e$ the number of edges.

### 7.3.1. Potential difference space $\delta \mathcal{P}$ and Bond space $\mathcal{B}$ (cut-set space)

Let $\delta \mathcal{P}$ be the space of all possible potential differences on the graph $G$. Let $\delta p, \delta q \in \delta \mathcal{P}$ be two potential differences functions and $\lambda \in \mathbb{R}$ a scalar. It can be seen that $\delta \mathcal{P}$ is closed under addition and multiplication by a scalar ($\delta p + \lambda \delta q \in \delta \mathcal{P}$). Consequently, the set of all possible potential differences on the graph constitutes a vector space. In this paper, potential differences associated with cut-sets are of special interest. Let $\Delta = [S, \bar{S}]$ be a cut-set, then the potential difference associated to $\Delta$ is denoted $\delta p^\Delta$ and is defined as:

$$\delta p^\Delta((i,j)) = \begin{cases} 1 & \text{if } (i,j) \in \Delta \text{ and same orientation} \\ -1 & \text{if } (i,j) \in \Delta \text{ and opposite orientation} \\ 0 & \text{if } (i,j) \notin \Delta \end{cases} \quad (A.1)$$

In the rest of the appendix, elementary cut-sets refer to cut-sets that isolate only one vertex $v$ and the associated potential difference will be denoted $\delta p^v$. In an elementary cut-set, $\delta p^v$ can be simply expressed as:

$$\delta p^v((i,j)) = \delta_{iv} - \delta_{jv} \quad (A.2)$$

where $\delta_{ij}$ is the Kronecker delta. A cut-set $\Delta$ can be described by a $1 \times n_e$ vector $b^\Delta$ with components in $\{-1, 0, 1\}$. If $b^\Delta$ is indexed by the set of edges, $E$, then $b^\Delta$ can be expressed as:

$$b^\Delta_{(i,j)} = \begin{cases} 1 & \text{if } (i,j) \in \Delta \text{ and same orientation} \\ -1 & \text{if } (i,j) \in \Delta \text{ and opposite orientation} \\ 0 & \text{if } (i,j) \notin \Delta \end{cases} \quad (A.3)$$

Consequently, the cut-set vector $b^{\Delta_1}$ corresponding to $\Delta_1$ in Figure 1 is $[0, -1, -1, 1, 0, 0, 1, 0, 0, -1]$ and $b^7$ corresponding to $\Delta_2 = \{\{7\}, V \setminus \{7\}\}$ is $[0, 0, 0, 0, 0, 0, 1, 0, 1, -1]$. One can see that the cut-set vector is another way of expressing the potential difference function as a row vector whose coordinates are indexed with edges in $E$. The components of this vector represent the values that the potential difference function takes on each edge $(i,j)$ with respect to a cut $\Delta$. The collection of all elementary cut-set row vectors corresponding to all the nodes in $V$ composes the matrix $A$ referred to in graph theory as the node-to-edge incidence matrix of the graph.

Let $\delta p$ be a potential difference function and $p^v$ the associated potential at the node $v$. Let $\delta p^v$ be the elementary potential difference functions associated to the node $v$. Then for any $(i,j) \in E$:

$$\sum_{v \in V} p^v \, \delta p^v((i,j)) = \sum_{v \in V} p^v (\delta_{iv} - \delta_{jv})$$
$$= \sum_{v \in V} p^v \delta_{iv} - \sum_v p^v \delta_{jv} \qquad (A.4)$$
$$= p^i - p^j = \delta p((i,j))$$

Equation 4 proves that any potential difference can be expressed as a linear combination of the elementary potential differences associated with every node in the graph. Conversely, every linear combination of the elementary potential differences is also a potential difference. In fact, if $\sum_{v \in V} \alpha_v \delta p^v$ is a linear combination of the elementary potential difference states associated to every node, then the function $p$ on $V$ where $p(v) = \alpha_v$ is a potential since $\delta p$ obeys a circuit law. Consequently, the potential difference space $\delta\mathcal{P} = span((\delta p^v)_{v \in V})$ and the associated bond space $\mathcal{B}$ correspond to the row space of the node-to-edge incidence matrix $A$. The set of the linearly independent elementary potential differences constitute thus a vector basis for $\delta\mathcal{P}$ and they are referred to as the potential difference modes.

### 7.3.2. The flow space $\mathcal{F}$ and the circuit space $\mathcal{C}$

Let $\mathcal{F}$ be the space of all possible flow functions on the graph $G$. Let $f, g \in \mathcal{F}$ be two flow functions and $\lambda \in \mathbb{R}$. The linear combination $(f + \lambda g)$ obeys a cut law on the graph $G$ proving that $\mathcal{F}$ is a vector space. Like potential differences, the flow functions associated to circuits are of interest for the description of the network. Let $C$ be a circuit in the graph $G$. The flow $f_C$ associated to the circuit $C$ is defined by:

$$f_C((i,j)) = \begin{cases} 1 & if \quad (i,j) \in C \\ -1 & if \quad (j,i) \in C \\ 0 & if \quad (i,j) \notin C \end{cases} \qquad (A.5)$$

Elementary circuits represent cycles in graph theory: closed paths where every vertex has exactly two neighbors. The space of all possible circuits is referred to as the circuit space $\mathcal{C}$ in graph theory, or cycle space $\mathcal{C}$ in some other works, which corresponds to the nullspace of the node-to-edge incidence matrix $A$. The change in the denomination of the space can be justified by the fact that the circuit space $\mathcal{C}$ can be studied through the set of independent cycles of the graphs. In fact, similar to cut-set vectors in the previous sections, every cycle $C$ corresponds to a $1 \times n_e$ vector $c^C$ with components in $\{-1,0,1\}$. The vector $c^C$ is indexed by the set of edges $E$ and is defined by:

$$c^C_{(i,j)} = \begin{cases} 1 & if \quad (i,j) \in C \text{ and same orientation} \\ -1 & if \quad (i,j) \in C \text{ and opposite orientation} \\ 0 & if \quad (i,j) \notin C \end{cases} \quad (A.6)$$

There is thus a direct association of the flow function $f_C$ to the circuit vector $c^C$. The cycle space $C$ has a finite dimension $s$ and its bases correspond to $s$ cycle vectors, referred to as the flow modes. The determination of these vectors is discussed in the next section, through its relationship with the cut-set space $B$ and the node-to-edge incidence matrix of the graph $A$.

### 7.3.3. Relationship between the potential space $\delta P$ and the flow space $F$

Let $f$ be a function on the edges $E$ of the graph $G$. $f$ is a flow if, and only, if it obeys a cut law on the graph $G$. Since the set of elementary cuts $(b^v)_{v \in V}$ spans the cut-set space $B$, $f$ has to obey a cut law on the cut-sets $(b^v)_{v \in V}$ in order to be a flow. These conditions can be expressed as:

$$\sum_{(i,j) \in E} b^v f\big((i,j)\big) = 0, \forall\, v \in V \quad (A.7)$$

Since each $b^v$ is associated to an elementary potential difference $\delta p^v$, it can be seen that $f$ is a flow if, and only if, it is orthogonal to each elementary potential difference $\delta p^v$, with the flow space $F$ representing the orthogonal complement of the potential difference space $\delta P$. Moreover, $F$ corresponds also to the nullspace of the node-to-edge incidence matrix $A$. Algebraically, the flow modes can be determined through the basis of the nullspace of the incidence matrix A. However, the resulting flow modes are typically associated with complicated circuits in the graph when higher dimensions are considered. Aloui et al. (2019) [A.2-A.4] proposed a bio-inspired method for the study and design of self-equilibrated networks in two- and three-dimensional spaces in the context of tensegrity structures. In the proposed method, called cellular morphogenesis, flow modes are referred to as self-stress states corresponding to tensegrity cells (complete graphs on $d + 2$ nodes) composing the overall structure. An algorithm was also provided for the determination of the tensegrity cells composing a given tensegrity structure, which is used in this paper for flow mode determination.

**Appendix B**

The expression of the analytical solution for the flow mode of a cell resorts to algebraic geometry and the wedge product or exterior product ∧. The reader is reminded here of some of the properties of wedge products used in this proof:

- The wedge product, or exterior product, is an algebraic construction used in algebraic geometry to study areas, volumes and higher-dimensional analogues.
- The wedge product of two vectors $u$ and $v$ is called a bivector $u \wedge v$, and it lives in a vector space called the exterior square distinct from the original vector space.
- The wedge product $u \wedge u = 0$. In general, the wedge product of linearly dependent vectors is always zero.

- The wedge product is anticommutative $u \wedge v = -v \wedge u$.
- The wedge product is associative $(u \wedge v) \wedge w = u \wedge (v \wedge w) = u \wedge v \wedge w$.
- $u \wedge v$ is also called a 2-blade and it represents the oriented area of the parallelogram defined by $u$ and $v$ and given the orientation described by Figure B.1.

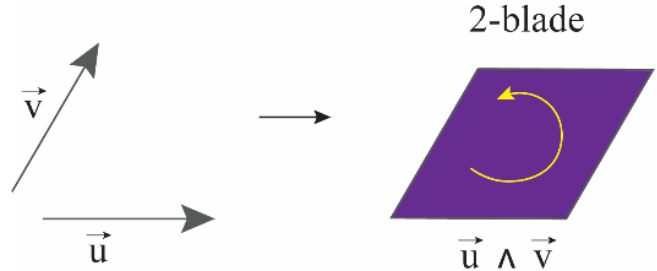

Figure B.1: Illustration of a 2-blade and its orientation.

- A k-blade $u_1 \wedge u_2 \wedge \ldots \wedge u_k$ is a generalization of the 2-blade associated to a $k$-dimensional oriented volume of a parallelotope. Figure B.2 shows an example of a 3-dimensional blade and its corresponding volume.

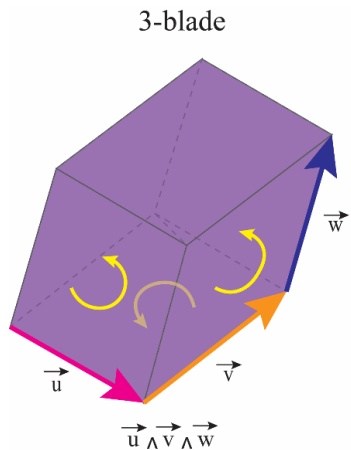

Figure B.2: Illustration of a 2-blade and the orientation of the volume it defines.

- It is common to consider a k-simplex instead of a parallelotope when the k-blade vectors are defined by two points in space. The $k$-simplex and its orientation can then be entirely described by an ordered set $(P_i)_{0 \leq i \leq k}$ of $k + 1$ points. The oriented volume can then be associated to the convex hull of the $k + 1$ nodes (the smallest convex set containing the $k + 1$ nodes of the k-simplex) endowed with the orientation of the ordered set of vertices. Consequently, the volume of the $k$-simplex defined by the ordered set $\{P_0, P_1, \ldots, P_k\}$ is described using $k$-blades by [B.1]:

$$V_{k-simplex} = \frac{1}{k!} \left( \overrightarrow{P_0 P_1} \wedge \overrightarrow{P_0 P_2} \wedge \ldots \wedge \overrightarrow{P_0 P_k} \right) \qquad (B.1)$$

- In a $d$-dimensional space with an orthonormal basis $(\vec{x_1}, \vec{x_2}, ..., \vec{x_d})$, the analytical value of the $d$-blade defined by $(\overrightarrow{P_0P_1} \wedge \overrightarrow{P_0P_2} \wedge ... \wedge \overrightarrow{P_0P_d})$ can be calculated by the determinant[B.1, B.2]:

$$\left(\overrightarrow{P_0P_1} \wedge \overrightarrow{P_0P_2} \wedge ... \wedge \overrightarrow{P_0P_k}\right) = \begin{vmatrix} 1 & P_{01} & \cdots & P_{d0} \\ 1 & P_{02} & \cdots & P_{d1} \\ \vdots & \vdots & \vdots & \vdots \\ 1 & P_{0d} & \cdots & P_{dd} \end{vmatrix} \quad (B.2)$$

In this paper, the wedge product is used to derive the analytical solution for the flow densities of the flow mode of a $d$-dimensional cell. Let $G(V, E)$ be the complete graph on $d + 2$ nodes associated to a $d$-dimensional cell. Let the set of ordered nodes $(v_i)_{1 \leq i \leq d+2}$ be the composing nodes of the graph $G$. The graph $G$ is embedded in a $d$-dimensional space where to each node $v_i$ is associated a $d$-dimensional potential $p_i$. In a complete graph on $d + 2$ nodes, each vertex is adjacent to exactly $d + 1$ nodes. The elementary cut law $b^{v_k}$ around a node $v_k$ chosen arbitrarily from $V$ will give:

$$\sum_{\substack{1 \leq i \leq d+2 \\ i \neq k}} \omega_{ki} \overrightarrow{p_k p_i} = 0 \quad (B.3)$$

Note that Eq (B.3) describes the nodal equilibrium at node $k$ and the vector $\overrightarrow{p_k p_i}$ represents the potential difference between nodes $v_i$ and $v_k$. Since the flow space $\mathcal{F}$ of a cell is one dimensional, it follows that all the flow densities of the members can be written as a function of one specific flow density, without loss of generality $\omega_{km}$ ($m \neq k$). The objective of this part is to write the flow density of an edge $(v_k, v_n)$ ($n \neq m \neq k$), $\omega_{kn}$, as a function of the flow density of the edge $(v_k, v_m)$, $\omega_{km}$. Consider the $(d-1)$-blade composed of the vectors $\overrightarrow{p_k p_i}$ where $1 \leq i \leq d+2$, $i \neq k \neq m \neq n$, and denoted:

$$\bigwedge_{\substack{1 \leq i \leq d+2 \\ i \neq k \neq m \neq n}} \overrightarrow{p_k p_i} \quad (B.4)$$

Note that this wedge product does not contain vectors $\overrightarrow{p_k p_m}$ and $\overrightarrow{p_k p_n}$. Applying the wedge product of this $(d-1)$-blade to equation B.3 gives:

$$\left(\sum_{\substack{1 \leq i \leq d+2 \\ i \neq k}} \omega_{ki} \overrightarrow{p_k p_i}\right) \wedge \left(\bigwedge_{\substack{1 \leq i \leq d+2 \\ i \neq k \neq m \neq n}} \overrightarrow{p_k p_i}\right) = 0 \quad (B.5)$$

Given that the wedge product of linearly dependent vectors is zero ($u \wedge u = 0$), the only terms that are left in the sum are those with $\omega_{km}$ and $\omega_{kn}$ (all other products will have at least one repeated vector):

$$\omega_{km}\left(\overrightarrow{p_k p_m} \wedge \bigwedge_{\substack{1 \leq i \leq d+2 \\ i \neq k \neq m \neq n}} \overrightarrow{p_k p_i}\right) + \omega_{kn}\left(\overrightarrow{p_k p_n} \wedge \bigwedge_{\substack{1 \leq i \leq d+2 \\ i \neq k \neq m \neq n}} \overrightarrow{p_k p_i}\right) = 0 \quad (B.6)$$

If $(p_i)_{1 \leq i \leq d+2}$ are linearly independent, then $\omega_{kn}$ can be easily expressed in terms of $\omega_{km}$. To simplify the notations, the wedge product in equation A.6 will be denoted by:

$$\left(\overrightarrow{p_k p_n} \wedge \bigwedge_{\substack{1 \leq i \leq d+2 \\ i \neq k \neq m \neq n}} \overrightarrow{p_k p_i}\right) = V\left(v_k, v_n, (v_i)_{\substack{1 \leq i \leq d+2 \\ i \neq k \neq m \neq n}}\right)\left(\bigwedge_{i=1}^{d+2} \overrightarrow{x_i}\right) \quad (B.7)$$

The function $V$ is synonymous with the determinant in equation B.2. Consequently, the order of vertices inside the function has to respect the same order of appearance of indices in the wedge product. Since the set $(v_i)_{1 \leq i \leq d+2}$ is ordered, then the set of ordered vertices $\left\{v_k, v_m, (v_i)_{\substack{1 \leq i \leq d+2 \\ i \neq m \neq n}}\right\}$ describes a set of vertices starting with vertex $v_k$, then vertex $v_m$, then the rest of the vertices following the same order of appearance in the ordered set $(v_i)_{1 \leq i \leq d+2}$. Consequently, equation B.6 becomes:

$$\begin{aligned}
& \omega_{km} V\left(v_k, v_m, (v_i)_{\substack{1 \leq i \leq d+2 \\ i \neq k \neq m \neq n}}\right)\left(\bigwedge_{i=1}^{d+2} \overrightarrow{x_i}\right) + \omega_{kn} V\left(v_k, v_n, (v_i)_{\substack{1 \leq i \leq d+2 \\ i \neq k \neq m \neq n}}\right)\left(\bigwedge_{i=1}^{d+2} \overrightarrow{x_i}\right) = 0 \\
\Rightarrow & \left[\omega_{km} V\left(v_k, v_m, (v_i)_{\substack{1 \leq i \leq d+2 \\ i \neq k \neq m \neq n}}\right) + \omega_{kn} V\left(v_k, v_n, (v_i)_{\substack{1 \leq i \leq d+2 \\ i \neq k \neq m \neq n}}\right)\right]\left(\bigwedge_{i=1}^{d+2} \overrightarrow{x_i}\right) = 0 \\
\Rightarrow & \omega_{km} V\left(v_k, v_m, (v_i)_{\substack{1 \leq i \leq d+2 \\ i \neq k \neq m \neq n}}\right) + \omega_{kn} V\left(v_k, v_n, (v_i)_{\substack{1 \leq i \leq d+2 \\ i \neq k \neq m \neq n}}\right) = 0
\end{aligned} \quad (B.8)$$

If the potentials $p_i$, $\forall\, i$, are linearly independent, then function $V$ cannot be zero, leading to:

$$\omega_{kn} = -\frac{V\left(v_k, v_m, (v_i)_{\substack{1 \leq i \leq d+2 \\ i \neq k \neq m \neq n}}\right)}{V\left(v_k, v_n, (v_i)_{\substack{1 \leq i \leq d+2 \\ i \neq k \neq m \neq n}}\right)} \omega_{km} \quad (B.9)$$

The examination of the two- and three-dimensional cases of Equation B.9 shows that there is another way of expressing flow density. Taking flow densities variables as constants of multiplication:

$$\omega_{km} = V\left(v_k, (v_i)_{\substack{1 \leq i \leq d+2 \\ i \neq k \neq m}}\right) \cdot V\left(v_m, (v_i)_{\substack{1 \leq i \leq d+2 \\ i \neq k \neq m}}\right) \quad (B.10)$$

and replacing $\omega_{km}$ in equation B.8 by it is expression in equation B.10 gives:

$$V\left(v_k, (v_i)_{\substack{1\leq i\leq d+2 \\ i\neq k\neq m}}\right) . V\left(v_m, (v_i)_{\substack{1\leq i\leq d+2 \\ i\neq k\neq m}}\right) . V\left(v_k, v_m, (v_i)_{\substack{1\leq i\leq d+2 \\ i\neq k\neq m\neq n}}\right)$$
$$+ \omega_{kn} V\left(v_k, v_n, (v_i)_{\substack{1\leq i\leq d+2 \\ i\neq k\neq m\neq n}}\right) = 0 \quad (B.11)$$

Now, without loss of generality, assume that $m < n < k$. The ordered set of vertices inside each function $V$ are reordered taking into consideration the changes to the values of $V$. Since $V$ is a determinant, every permutation of a vertex will result in multiplying $V$ by -1:

$$V\left(v_k, (v_i)_{\substack{1\leq i\leq d+2 \\ i\neq k\neq m}}\right) = V\left(v_k, (v_i)_{\substack{1\leq i\leq d+2 \\ i\neq k\neq m}}\right) \text{ (i)}$$
$$V\left(v_m, (v_i)_{\substack{1\leq i\leq d+2 \\ i\neq k\neq m}}\right) = (-1)^{m+n-2} V\left(v_n, (v_i)_{\substack{1\leq i\leq d+2 \\ i\neq k\neq n}}\right) \text{ (ii)}$$
$$V\left(v_k, v_m, (v_i)_{\substack{1\leq i\leq d+2 \\ i\neq k\neq m\neq n}}\right) = (-1)^{m-1} V\left(v_k, (v_i)_{\substack{1\leq i\leq d+2 \\ i\neq k\neq n}}\right) \text{ (iii)}$$
$$V\left(v_k, v_n, (v_i)_{\substack{1\leq i\leq d+2 \\ i\neq k\neq m\neq n}}\right) = (-1)^{n-2} V\left(v_k, (v_i)_{\substack{1\leq i\leq d+2 \\ i\neq k\neq m}}\right) \text{ (iv)}$$
$$\quad (B.12)$$

In equation B.12 (i), there was no need to reorder the vertices. In equation B.12 (ii), node $v_n$ is changed to the first position of the ordered set and node $v_m$ is put back inside the remaining of the set in its order of appearance in the original set $(v_i)_{1\leq i\leq d+2}$. In equation B.12 (iii), $v_m$ is put back inside the remaining of the set in its order of appearance in the original set $(v_i)_{1\leq i\leq d+2}$. In equation B.12 (iv), $v_n$ is put back inside the remaining of the set in its order of appearance in the original set $(v_i)_{1\leq i\leq d+2}$. Now, with these changes, equation B.11 can be rewritten as:

$$V\left(v_k, (v_i)_{\substack{1\leq i\leq d+2 \\ i\neq k\neq m}}\right) . (-1)^{m+n-2} V\left(v_n, (v_i)_{\substack{1\leq i\leq d+2 \\ i\neq k\neq n}}\right) . (-1)^{m-1} V\left(v_k, (v_i)_{\substack{1\leq i\leq d+2 \\ i\neq k\neq n}}\right)$$
$$+ \omega_{kn} (-1)^{n-2} V\left(v_k, (v_i)_{\substack{1\leq i\leq d+2 \\ i\neq k\neq m}}\right) = 0$$
$$\Rightarrow V\left(v_k, (v_i)_{\substack{1\leq i\leq d+2 \\ i\neq k\neq m}}\right) \left[V\left(v_n, (v_i)_{\substack{1\leq i\leq d+2 \\ i\neq k\neq n}}\right) V\left(v_k, (v_i)_{\substack{1\leq i\leq d+2 \\ i\neq k\neq n}}\right) - \omega_{kn}\right] = 0$$
$$\omega_{kn} = V\left(v_k, (v_i)_{\substack{1\leq i\leq d+2 \\ i\neq k\neq n}}\right) V\left(v_n, (v_i)_{\substack{1\leq i\leq d+2 \\ i\neq k\neq n}}\right)$$
$$\quad (B.13)$$

which gives $\omega_{kn}$ as:

$$\omega_{kn} = V\left(v_k, (v_i)_{\substack{1\leq i\leq d+2 \\ i\neq k\neq n}}\right) V\left(v_n, (v_i)_{\substack{1\leq i\leq d+2 \\ i\neq k\neq n}}\right) \quad (B.14)$$

The consideration of all possible rearrangements of $k, m$ and $n$ leads thus to the same result. Given that $k, m$ and $n$ were chosen arbitrarily, this expression is valid for the flow density of any edge in the cell. Moreover, the geometric interpretation of this result reflects that each flow density $\omega_{ij}$ of an edge $(v_i, v_j)$ of a $d$-dimensional cell can be calculated by the product of two oriented volumes of $d$-simplexes $S_i$ and $S_j$ defined by the oriented set of nodes $S_i = \left\{v_i, (v_k)_{\substack{1 \leq k \leq d+2 \\ k \neq i \neq j}}\right\}$ and $S_j = \left\{v_j, (v_k)_{\substack{1 \leq k \leq d+2 \\ k \neq i \neq j}}\right\}$. Figure B.3 illustrates the geometric interpretation of the result in two- and three-dimensions.

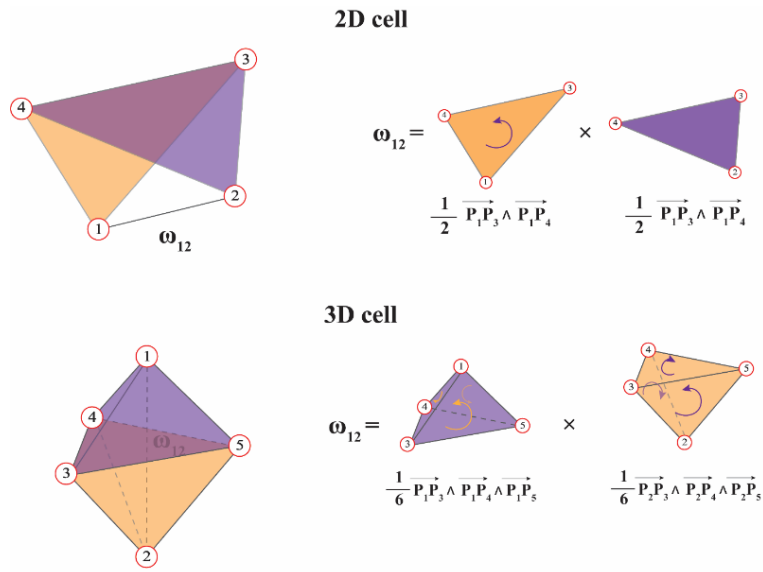

*Figure B.3: Illustration of the geometric interpretation of the flow mode solutions in two- and three-dimensions.*